\documentclass{aa}  

\usepackage{graphicx}
\usepackage{array}    
\usepackage{amssymb}    
\usepackage{url}
\usepackage{color}
\usepackage{multirow}
\usepackage{makecell}
\usepackage{adjustbox}
\usepackage{txfonts}
\usepackage{multicol}
\usepackage{longtable}

\begin{document} 

   \title{On the origin of sinusoidal brightness variations in F to O-type stars through radial velocities}

   \author{E. \v{S}ipkov\'a
          \inst{1,2}         
          \and
          M. Skarka\inst{3}
          \and
          M. Va\v{n}ko\inst{4}
          \and
          V. Chmela\v{r}\inst{1}
          \and
          T. Pribulla\inst{4}
          \and
          Z. Mikul\'{a}\v{s}ek\inst{1}
          }
   \institute{Department of Theoretical Physics and Astrophysics, Masaryk University, Kotl\'{a}\v{r}sk\'{a} 2, CZ-61137 Brno, Czech Republic
        \and
            Institute of Astronomy, KU Leuven, Celestijnenlaan 200D, 3001 Leuven, Belgium
        \and
            Astronomical Institute of the Czech Academy of Sciences, Fri\v{c}ova 298, CZ-25165 Ond\v{r}ejov, Czech Republic
        \and
            Astronomical Institute of the Slovak Academy of Sciences, 059 60 Tatransk\'{a} Lomnica, Slovak Republic
             }

   \date{Received October xx, 2025; accepted October xx, 2025}

  \abstract
   {
   Stellar variability may originate from various phenomena such as binarity, pulsations, or rotation. These mechanisms can induce flux variations of similar magnitudes, shapes, and periods.
   }
   {
   We aim to determine mechanisms responsible for the sinusoidal variations in main-sequence stars hotter than 6500\,K.
   }
   {
   We conducted our analysis using TESS long-cadence data complemented with high-resolution spectra from three spectrographs. From the initial sample of almost 46\,000 objects, we selected 35 targets for spectroscopic follow-up. Comparison of light curves and radial velocity curves allowed for robust classification of these targets.
   }
   {
   Among the 35 selected objects, 18 displayed variability, suggesting the presence of a companion (including the discovery of 7 new binary systems and 1 candidate for a triple-star system), 1 was identified as a new pulsator, 9 as new candidates for spotted stars, and 7 objects had uncertain classification. Our analysis shows that at least half of randomly selected stars with sinusoidal brightness variations are binaries.
   }
   {
   The presented results illustrate the need for an individual approach to stellar classification, especially in cases where the photometric data alone is insufficient for determining the underlying phenomena behind the observed variations.
   }

    \maketitle

\section{Introduction}\label{Sect:Introduction}
Various astrophysical processes can induce photometric variations, which may result in sudden, non-periodic, or quasi-periodic brightenings or periodic patterns. Periodic variations can be attributed to eclipses, pulsation, rotation, or other conditions and processes \citep[e.g.,][]{fetherolf}. Understanding their origin is crucial for obtaining information about the dynamics, internal structure, and other fundamental properties of these objects \citep{kurtz}.

There have been several attempts to classify variable objects into distinct categories based on shared properties, for example, an updated Pickering's classification \citep{townley} or classification in the General Catalog of Variable Stars \citep[GCVS,][]{samus}. Although such classifications are useful, transitions between classes can be continuous \citep{eyer}, and the properties may overlap between different categories, making it difficult to identify the source of variability without additional information (e.g., spectra, observations in different wavebands; \citealt{skarka22}).

This work utilises the classification first proposed by \citet{eyer} and revised in \citet{eyer2018}, which has been widely used in large-scale surveys, such as Gaia DR2 \citep{eyer2019} and WISE \citep{petrosky}.
This classification employs four levels of division. The first level distinguishes between the variability caused by intrinsic and extrinsic processes. The second level groups objects according to their type (asteroids, stars, AGN), which are further divided according to the mechanism responsible for the variability. At last, objects with similar photometric properties, such as period, amplitude, timescale, and the shape of the light curve, are gathered in the last level of division \citep{eyer}.

In our study, we focus on variable hot main-sequence stars with spectral classes F-O. The most massive and luminous stars are valuable probes of stellar formation and galactic structure due to their short lifespans, strong winds, and rapid evolution (see, for example, \citealt{zinnecker}; \citealt{marchant}; \citealt{eldridge}). Stars of F-O spectral types are also useful for our understanding of the internal structure of stars, mixing processes, rotation, and angular-momentum transfer by interpreting their oscillations \citep[see, for example, reviews by][]{Aerts2019,Aerts2021,kurtz,Aerts2024}. In general, hot stars\footnote{Stars hotter than about $\approx 6250$\,K \citep{kraft}.} can be characterised by radiative outer layers and absence of activity connected with convection, although some recent investigations show that there might be effects similar to those observed in cool stars present also in hot stars. For example, \citet{Henriksen2023} and \citet{Antoci2025} found that rotational modulation in some hot stars is likely induced by spots of higher or lower temperature and sub-surface convection.

The rotation of F-O stars spans from slow to fast rotators \citep[see, e.g.,][]{Royer2007}. In stars, where the mixing processes are negligible, typically stars with equatorial rotation velocity less than about 100\,km/s \citep{Mathys2004,Abt1995,Qin2021}, we often observe unusual chemical composition, thus, chemical peculiarity \citep[CP,][]{Maury1897,Preston1974,Schnell2008}. Currently, we know tens of thousands of CP stars of various types \citep[e.g.,][]{Ghazaryan2018,Lu2025}. It is generally accepted that the CP phenomenon is a product of atomic diffusion in calm atmospheres when elements with large cross-sections are elevated to the surface layers while elements with small cross-sections settle down \citep{Michaud1970,Michaud1976}.

These elements can create spots that may be long-living and stabilised by a strong, globally organised magnetic field, forming a class of magnetic chemically peculiar stars, shortly mCP or Ap/Bp stars \citep{Preston1974,Alecian2015}. The mCP stars are characterised by the overabundance of Si, Cr, Fe, Sr, and rare earth elements \citep{Preston1974}. The magnetic fields in mCP stars can be very strong, even in the order of tens of kG \citep{Babcock1960,Bagnulo2003,Shultz2019} and stable over decades \citep[e.g.,][]{Donati2009,Oksala2012,Shultz2018}. The incidence rate of mCP stars among F-O stars is less than about 10\,\% \citep{Donati2009,Sikora2019b}. The origin of the magnetic field in hot stars without outer convective zones is assumed to be of a fossil nature surviving from the time of their formation via cloud collapse, or mergers \citep[see, e.g.,][]{Braithwaite2004,Ferrario2018,Schneider2019}. The impact of strong magnetic fields on stellar evolution, rotation, diffusion processes, and mass loss is still not fully understood \citep[see e.g.,][]{Shultz2019}.

Because mCP stars usually show stable rotational modulation that is easily detectable from photometric data, they are usually searched via this channel \citep[e.g.,][]{Hummerich2016,Bernhard2021,Sikora2019,Barron2020}. However, close non-eclipsing binary stars called ellipsoidal variables, \citep[ELL,][]{Beech1985,Morris1985} can show similar light curves with similar amplitudes and periods due to gravity darkening of their tidally-deformed components, Doppler beaming, and reflection effect \citep[e.g.,][]{Faigler2012,Green2023}. As was discussed by \citet{skarka22}, it is not possible to distinguish between a spotted rotationally variable star and an ELL star from single-channel photometry. \citet{Faigler2012} gave examples of seven ELL stars confirmed spectroscopically that show light curves perfectly mimicking rotational modulation seen in single spotted stars. {\citet{Green2023} found that only about 50\,\% of suspected 97 ELL variables show clear radial velocity variations that firmly confirm their variability, leaving a large space for misclassification. In addition, pulsations might mimic rotational variability, preventing proper classification and interpretation of the photometric data \citep[see the discussion in][]{skarka22,Skarka2024}.

Among O to F main-sequence stars, at least about 50\,\% of stars are variable \citep{eyer2019,skarka22,Skarka2024}. The subject of this paper is stars that show stable sinusoidal brightness variations, suggesting spots or ellipsoidal variability. Because it is impossible to distinguish spots from binarity based solely on the light-curve shape, our main goal is to find out which stars are bound in binary systems and identify candidates for spotted stars based on spectroscopy, particularly the radial velocity analysis. From an initial sample of almost 46\,000 objects in the TESS data \citep{ricker}, we defined a sample of more than a hundred bright stars showing sinusoidal brightness variations with the selection procedure described in Section 2. For 35 stars, we complemented the TESS light curves with spectroscopic data from three high-resolution spectrographs (Section~\ref{sec03}). We analysed each star individually employing these spectroscopic observations, photometric information from TESS, and results from Gaia (Section~\ref{Sect:Analysis}). We discuss and conclude our results in Sections~\ref{Sect:Discussion} and \ref{Sect:Conclusions}.

\section{Sample definition}
\label{sec02}
The sample of hot main-sequence stars used in this paper was based on the photometric observations from the TESS mission \citep{ricker}. Data products include Target Pixel Files (TPFs), Light Curve Files (LCFs), and Full-Frame Images (FFIs) processed by the TESS Science Processing Operations Centre \citep[SPOC;][]{Jenkins2016}.
The analysis was carried out on LCFs with the provenance name TESS-SPOC, as most stars lacked data with the SPOC designation. Because the Quick-Look Pipeline (QLP) removes large-amplitude trends \citep{huang1} and, in its first years, removed also stellar variability as noted by \citet{skarka22}, we opted for a more conservative approach and have not used the QLP data. Similarly to \citet{skarka22}, we applied no additional detrending, since the pre-search data conditioning simple aperture photometry (PDCSAP) fluxes were generally free of significant outliers and trends.

\subsection{Initial sample}

The initial sample of stars was constrained from the TESS Input Catalogue (TIC) v8.0 \citep{strassun} using two criteria. The first criterion restricted the sample to stars with variability associated with the radiative processes in their envelopes. Because there is no exact boundary between the hot stars with radiative envelopes and the cold stars with convective envelopes, we considered several lower limits of the effective temperatures ($T_\mathrm{eff}$): 6000~K as the minimum $T_\mathrm{eff}$ of F-type stars and 6700~K as the temperature that defines the transition between radiative and convective envelopes related to the Kraft break \citep{anders, brun}. To account for stars near the transition region and minimise contamination from stars whose variability is associated with convective processes, we adopted a value of 6500~K.
The second criterion limited the observed magnitude in the Johnson $V$ band to stars brighter than $V=9~\mathrm{mag}$. It was chosen for practical reasons - we selected only stars that were suitable as targets for high-resolution echelle spectra taken with 1-metre class telescopes. More details about the used instruments can be found in Section \ref{sec03}.

The initial sample contained 45780 stars. 
We collected available data from MAST using \textsc{Lightkurve} software \citep{Lightkurve2018, Barentsen2020}. For quick visual examination of the targets, we generated figures similar to those from \citet{skarka22} that contain key stellar characteristics. These images were generated in December 2023 using available data up to sector 72 (Cycle 6) of the TESS mission.

To analyse the periodicity of unevenly sampled data, we constructed Lomb-Scargle periodograms \citep{lomb, scargle} in a low-frequency regime ($0-5~\mathrm{c/d}$) and a high-frequency regime ($5-100~\mathrm{c/d}$) and included them in the visual representation images. The significance of the signal was assessed using the false alarm probability (FAP), which evaluates the likelihood that the peak arises from noise. We chose a 1\% FAP threshold.
The light curves were phased with both the dominant frequency from the low- and high-frequency regimes. The visual representation images also contained an image of the observed object as seen by the TESS cameras for examination of the stellar background and identification of blends.

To obtain a more manageable dataset, we reduced the initial sample to objects that exhibited simple sinusoidal variations in the low-frequency regime, where various mechanisms can explain the variability. Variations in the high-frequency regime of hot stars are usually associated with pulsations, and their classification was beyond the scope of this work. Additionally, variations in the high-frequency regime could originate from Nyquist reflections. We employed a careful examination of the observed variability to identify and discard aliases during the candidate selection process, so the sample contained only low-frequency sinusoidal signals. We inspected the phased low-frequency light curves for shape deformations and minima of various depths that deviated from the simple sinusoidal shape, discarding objects that did not adhere to our requirements. 

The reduced sample contained 472 objects. Nonetheless, visual inspection had its limitations and could not eliminate all non-sinusoidal light curves. These objects were carried forward with the understanding that a more thorough examination would be employed during finalisation of the sample in the next step.

\subsection{Finalisation of sample}

The presence of strictly sinusoidal variations was assessed by investigating the residuals after fitting a sinusoidal model to the photometric data. For each object from the reduced dataset, we selected one sector from the TESS observations least affected by instrumental deformations. The light curve was fitted with a simple sine function $y_\mathrm{SSF}$ (Eq. \ref{eq2sin}) and a function containing two sines $y_\mathrm{TSF}$ (Eq. \ref{eq2sin2}) with frequencies $f$ and $f/2$.
\begin{equation}
y_\mathrm{SSF} = a_1\sin \left( 2 \pi f t + \phi_1 \right)
\label{eq2sin}
\end{equation}
\begin{equation}
y_\mathrm{TSF} = a_1\sin \left( 2 \pi f t + \phi_1 \right) + a_2\sin \left( \pi f t + \phi_2 \right)
\label{eq2sin2}
\end{equation}

An example of a figure used for the evaluation of light curve shapes is shown in Fig.~\ref{fig:res}, where the blue and orange curves are model fits based on Eq.~\ref{eq2sin} and Eq.~\ref{eq2sin2}, respectively. We calculated the difference between the data and individual models, which are shown in the residuals plots in Fig.~\ref{fig:res} and evaluated the goodness of each fit by estimating their standard deviations (STDs).

\begin{figure*}
\centering
   \includegraphics[width=17cm]{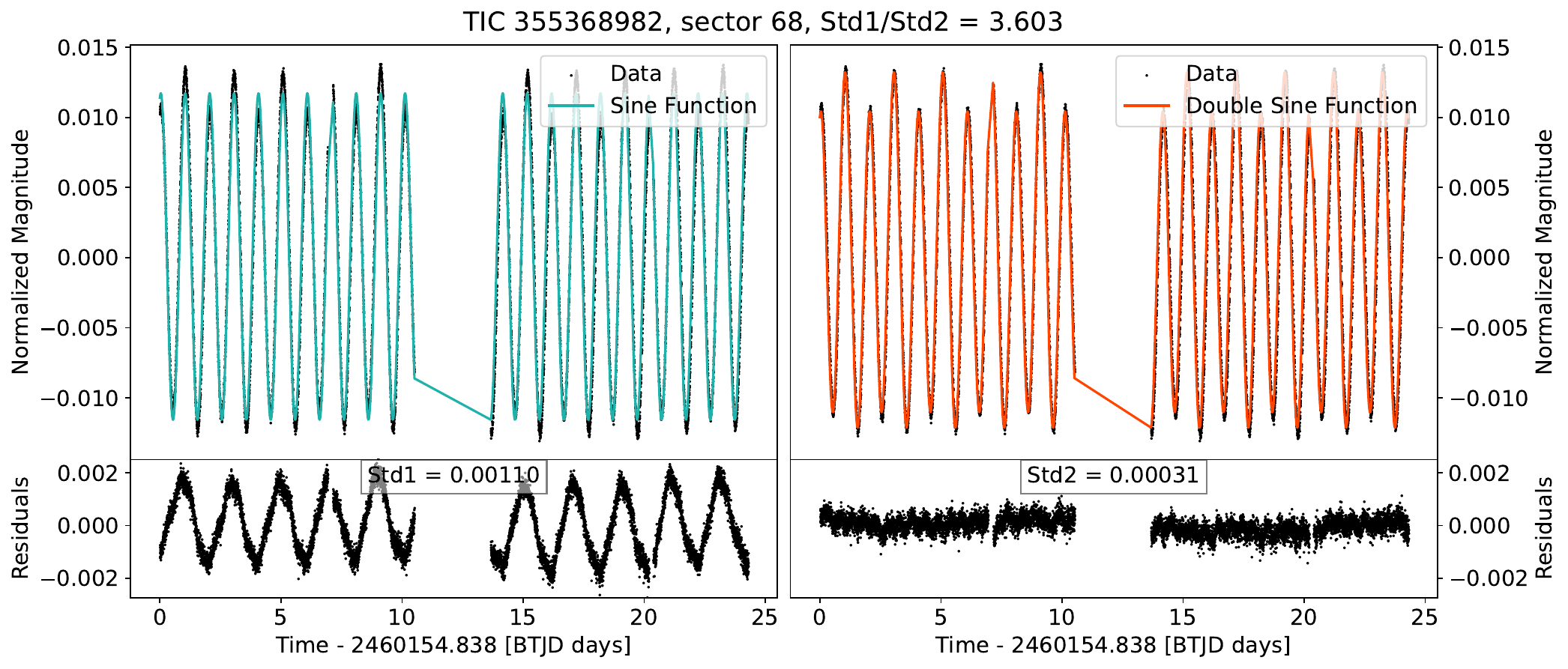}
     \caption{Residuals after fitting the data with a single-sine (left panel) and a two-sine (right-hand panel) function. The figure contains standard deviations of residuals and their ratio used for identifying sinusoidal variability.}
     \label{fig:res}
\end{figure*}

The fitted models were compared by calculating the ratio of STDs for the function with two sines and the single-sine function. We applied four criteria for the selection of sinusoidal variables from the reduced sample. First, the residuals ratio needed to be below a value of 1.1. A ratio close to 1 indicated an ideal case where both functions fit the data in the same manner; however, due to data scatter, the two-sine function fit the data better in almost all cases.  A ratio above 1.1 was empirically determined as the threshold where the two-sine function fit the data significantly better, indicating that two consecutive minima had different depths that could not be explained by data scatter. These objects were rejected from the final sample, as these variations indicated their binary nature.

The second criterion rejected stars that exhibited light curve modulation due to beating. The amplitude of variations periodically changes with time, as the star pulsates. As stated before, the classification of pulsation modes is beyond the scope of this work, and these objects were rejected. The third criterion discarded stars that changed the shape of their light curves or the period of variations, while the fourth rejected objects with visible periodic trends in their residuals caused by a non-sinusoidal shape. The number of stars eliminated using each criterion and the targets contained in the final sample can be seen in Table~\ref{tab2res}.

\begin{table}[ht!]
\centering
\caption{Distribution of stars rejected due to criteria imposed by the study of their residuals.}
\label{tab2res}
\begin{tabular}{lcc}
\hline \hline
Criteria &
\multicolumn{1}{c}{$N$} &
\multicolumn{1}{c}{\begin{tabular}{c} Percentage \\ {(}\%{)} \end{tabular}} \\
\hline
Residuals ratio above 1.1    & 171   & 36.2  \\
Light curve modulation     & 47    & 10.0  \\
Shape/period change        & 14    & 3.0   \\
Non-sinusoidal shape       & 132   & 27.9  \\ \hline
Final sample               & 108   & 22.9  \\ \hline \hline
\end{tabular}
\end{table}

The final sample of stars contained 108 objects that were divided into stars observed in the northern and southern hemisphere based on their declination angles. The distribution of spectral types in the final sample can be seen in Table~\ref{tab2final}.

\begin{table}[ht!]
\centering
\caption{Distribution of spectral types in the final sample of stars.}
\label{tab2final}
\begin{tabular}{c|cccc}
\hline \hline
\multicolumn{1}{c}{Type} &
\multicolumn{1}{c}{\begin{tabular}{c} $T_\mathrm{eff}$ \\ {(}K{)} \end{tabular}} &
\multicolumn{1}{c}{Position} &
\multicolumn{1}{c}{\begin{tabular}{c} $N$ \\ \end{tabular}} &
\multicolumn{1}{c}{\begin{tabular}{c} Percentage \\ {(}\%{)} \end{tabular}} \\
\hline
\multirow{2}{*}{O}                & \multirow{2}{*}{\textgreater 25000}   & north  & 0  & 0       \\
&        & south  & 0  & 0       \\ \hline
\multirow{2}{*}{B}               & \multirow{2}{*}{11000\,-\,25000}             & north  & 9  & 8.3    \\
&        & south  & 18 & 16.7   \\ \hline
\multirow{2}{*}{A}               & \multirow{2}{*}{7500\,-\,11000}              & north  & 25 & 23.2   \\
&        & south  & 31 & 28.7   \\ \hline
\multirow{2}{*}{F}                & \multirow{2}{*}{6500\,-\,7500}              & north  & 11 & 10.2   \\
&        & south  & 14 & 13.0   \\ \hline \hline
\end{tabular}
\end{table}

\section{Spectroscopic observations}\label{sec03}
We supplemented the photometric data with spectroscopic observations that provided further insight into the origin of the photometric variability. From the 108 objects in the final sample, 45 targets were available for spectroscopic observations in the north with Ond\v{r}ejov Echelle Spectrograph \citep[OES;][]{oes1, oes2} and MUSICOS \citep{musikos}, while 63 targets were available for southern observations with PUCHEROS+ \citep{Antonucci2025}.

OES is a high-resolution instrument mounted on the 2-metre Perek telescope at Ond\v{r}ejov Observatory. It covers the wavelength range 3870-9200\,\AA~with a spectral resolving power $R=51600$, centred at 5000\,\AA ~\citep{oes1}. The long-term radial-velocity stability is approximately 200-300~m/s \citep{oes2}. An iodine absorption cell can be used as a wavelength reference for more precise RV measurements \citep{Karjalainen2022}. The limiting observation magnitude in the {\it V} filter is around 13~mag \citep{oes2}.

MUSICOS is a high-dispersion echelle spectrograph mounted at the 1.3-metre telescope located at Skalnat\'{e} Pleso Observatory (altitude of 1786~m above sea level). The high altitude reduces atmospheric turbulence and lowers water vapour content, improving the quality of spectroscopic observations. MUSICOS covers a spectral range of 4250-7375~\AA~with spectral resolution $R=25000-38500$, depending on the focusing \citep{musikos}. \citet{musikos} reports the RV stability of $100-200$\,m/s.

Another instrument that benefited from its high-altitude location was PUCHEROS+ \citep{Antonucci2025}, located at the ESO La Silla observatory at 2400~m above sea level. Mounted on an ESO E152 1.52-metre telescope, PUCHEROS+ covers a spectral range between 4000-7300~\AA~with $R=18000$ and with an RV precision of 30~m/s.
It was installed as a part of the PLATO spectroscopic follow-up programme and was replaced by the PLATOSpec instrument in November 2024 \citep{Kabath2025}.

Observations lasted from May to September 2024 in the southern hemisphere and from April 2024 to July 2025 in the north. Observing all 108 targets was not feasible based on the available observation time, so we selected the best candidates for spectroscopic follow-up.

We used the Image Reduction and Analysis Facility \citep[\texttt{IRAF};][]{tody} for the data reduction and RV estimations. We performed standard procedures of bias and flat field corrections. For the removal of the cosmic hits, we used \texttt{Dcr} routine \citep{Pych2004}. The spectra have been calibrated using Thorium-Argon lamps and corrected with the heliocentric RV and JD corrections.
The RVs were determined relative to a template spectrum using the cross-correlation method implemented in the \texttt{Fxcor} task within \texttt{IRAF}. The resulting cross-correlation function (CCF) provided the radial velocity shift between the two spectra caused by the Doppler effect.
The reference spectrum was selected as the highest-quality spectrum out of the observations. For targets with multiple visible components, the spectra were compared with a single-star template of a similar spectral type from \citet{oestemp}. All steps in the data reduction were performed with custom semi-automated \texttt{IRAF} scripts.

\section{Analysis}\label{Sect:Analysis}

We examined the sinusoidal variability of 35 objects with spectroscopic follow-up. Photometric data were first examined in December 2023 during the construction of the initial sample; however, since then, TESS continued observations, providing new data for several targets. The extended time baseline was used for a more precise determination of the variability period. Frequency spectra were examined with a software package \texttt{Period04} \citep{lenz}, to identify high-frequency signals, model the light curves, and refine frequencies, amplitudes, and phases.

Using the new TESS data, the difference between the initial variability period estimate and the period derived with \texttt{Period04} was less than one minute for 18 targets. However, for 17 objects, discrepancies reached up to 21 minutes, potentially blurring, shifting, or distorting the phase-folded light curves and RV curves. In the rest of the analysis, we used period estimates from \texttt{Period04}.

The majority of the targets did not exhibit high-frequency signals, suggesting the variability was dominated by a single periodic component. Ten objects contained high-frequency signals with significantly lower amplitudes (around 18 times lower on average). In several cases, the secondary peaks were identified as harmonics (integer multiples of the dominant frequency) or known instrumental artifacts. For two objects - TIC~16878120 and TIC~88815918 - the secondary peaks were within the Rayleigh resolution limit of the dominant frequency, meaning the dominant signal was not fully removed during pre-whitening. This signal could be explained by secular variations of the period caused by a variety of mechanisms, including the evolution processes of pulsators and dynamic interactions in binaries. TIC~320692159, on the other hand, displayed high-frequency peaks typical of $\delta$-Scuti pulsators \citep[DCST; see, for example, ][]{breger1,breger2,skarka22}.

The classification criteria were based on the combination of light curves (LCs; see, for example, the top plot of Fig.~\ref{fig:sb2}) obtained from photometric data and radial velocity (RV) curves from spectra (see, for example, the bottom plot of Fig.~\ref{fig:sb2}). We examined LCs and RVs phase-folded with the dominant period $P$ and twice the dominant period $2P$ (see, for example, the first and the second column of Fig.~\ref{fig:sb1} respectively). The spectral features (such as hydrogen lines), and CCFs (see Fig.~\ref{fig:ccf}) were used to support the classification. In cases with low signal-to-noise ratio (SNR) where the RV fit failed to converge, they also provided evidence based on identifiable features. The detailed description of the classification of individual variability types is provided in the following sections.
Each object was cross-referenced with the Simbad database \citep{wenger}, VSX catalogue \citep{vsx} and catalogue of chemically peculiar stars \citep{Renson2009}. We found 15 objects to be known as variable sources in Simbad, twelve objects were known as variable stars and present in the VSX, while seven stars were marked as chemically peculiar in \citet{Renson2009}. We note that detailed investigation of stars/systems, their chemical abundances, and modelling of the systems is beyond the scope of the current study.

\subsection{Binary stars}
Objects in the sample had low amplitudes and short periods, which, in the case of binary stars, can be explained by ellipsoidal variability. At least one of the components has a gravitationally distorted shape that causes brightness variations with two similar minima depths per cycle. The true photometric period is twice the observed one; therefore, it is necessary to examine data phase-folded with $2P$. The non-sinusoidal shape of the RV curve can be attributed to systems with eccentric orbits, Doppler beaming (change in observed flux due to the motion of a component towards or away from the observer), star spots, or reflection effect (the mutual irradiation of the components). Binary stars were distinguished from the sample in several ways, each of which can be explained by the orbital motion of the components around a common centre of mass. 

First, the spectral lines of both components were visible, leading to two separate RVs in anti-phase, as seen in the bottom plot of Fig.~\ref{fig:sb2}. The true orbital period of these components was $2P$, and the LCs and RVs remained correlated, with no RV shift (with respect to the template) during the minimum brightness, when the semi-major axis of the binary system was aligned with the observer's line of sight, and we observed no Doppler shift. The multiplicity of the signal could be verified by the CCFs, as seen in Fig.~\ref{fig:ccf}. These objects were marked as double-lined spectroscopic binaries (SB2).

\begin{figure}
    \centering
    \includegraphics[width=0.93\linewidth]{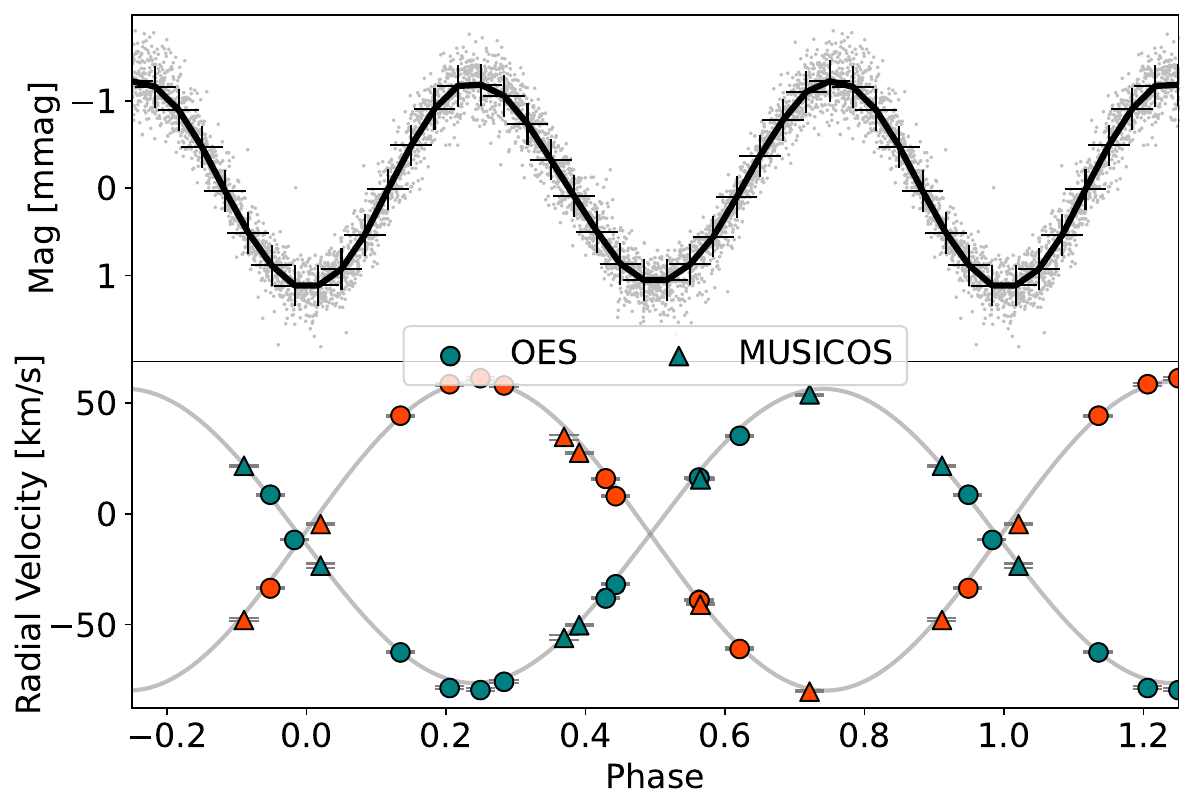}
    \caption{Example of LC (top plot) showing original (gray dots) and binned data (black crosses) and RV curve (bottom plot) for a spectroscopic binary with two visible components - TIC~257456854.}
    \label{fig:sb2}
\end{figure}

\begin{figure}
    \centering
    \includegraphics[width=0.93\linewidth]{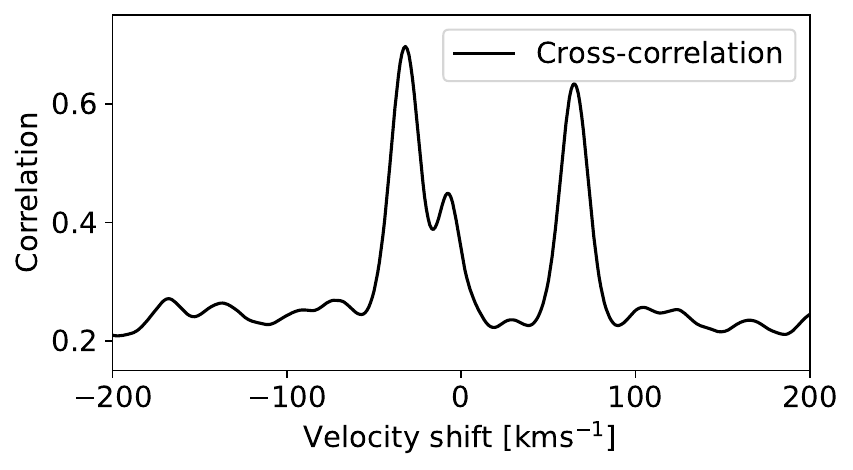}
    \caption{CCF of TIC~14400891. This object is a candidate for a triple star system, based on the number of correlation peaks.} 
    \label{fig:ccf}
\end{figure}

In the second case, we observe only one component of the binary system, marked as a single-line spectroscopic binary (SB1; see Fig.~\ref{fig:sb1}). When the data are phase-folded with period $P$, we observe weak or no variations in RVs. However, when folded with $2P$, the RV curve reveals a clear monotonic variation (see the bottom plots of Fig.~\ref{fig:sb1}). Additionally, the bottom left plot in Fig.~\ref{fig:sb1} mimics the variations caused by SB2, even though the measured spectra had only one visible component. This effect can be explained by phase folding an RV curve with half the correct period, and should not be confused with variations caused by SB2.

\begin{figure}
    \centering
    \includegraphics[width=0.93\linewidth]{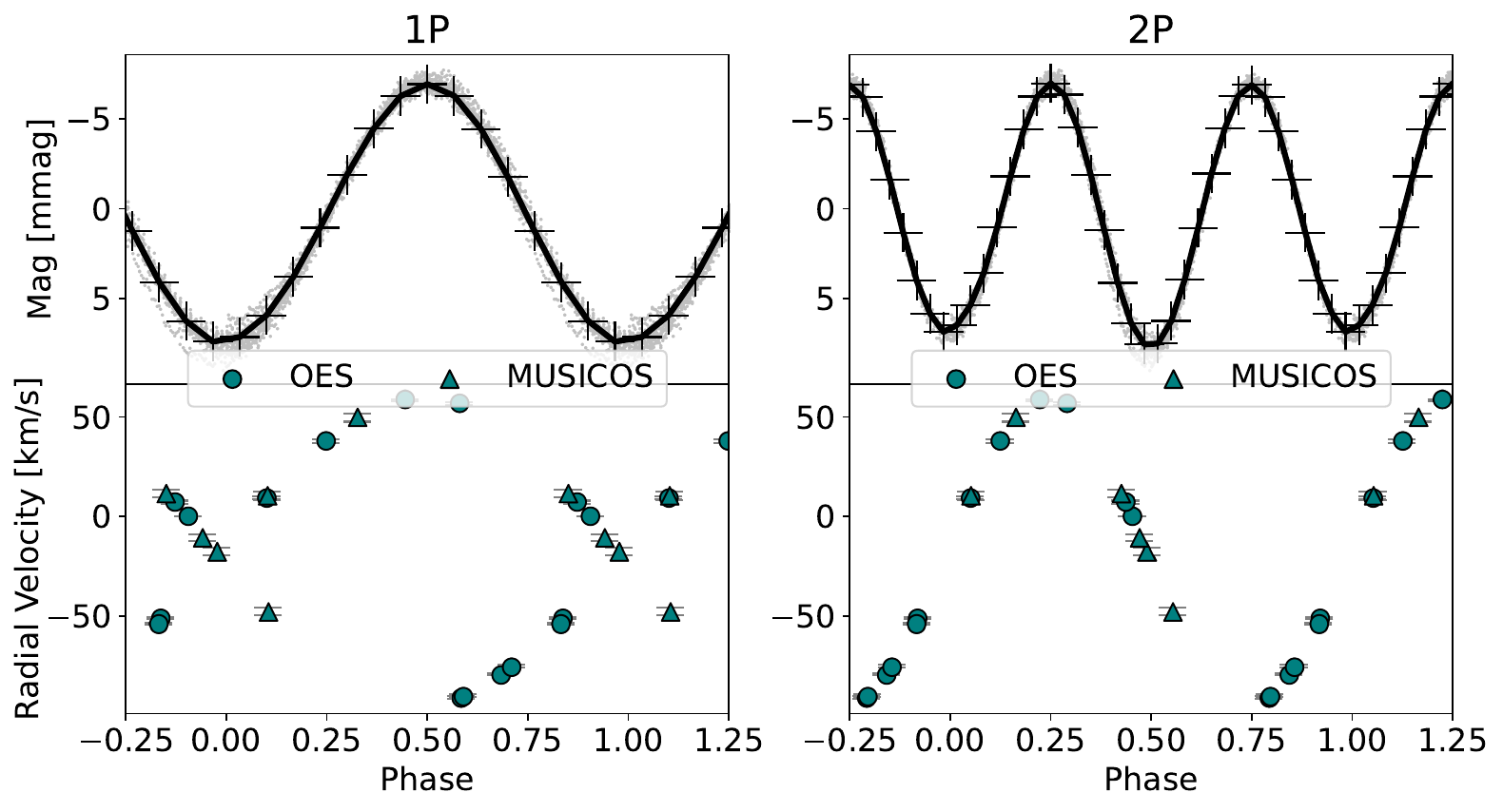}
    \caption{Comparison of LC (top plots) showing original (gray dots) and binned data (black crosses) and RVs (bottom plots) for a spectroscopic binary with one visible component phased with 1P (first column) and 2P (second column) - TIC~21673730.}
    \label{fig:sb1}
\end{figure}

Lastly, in cases where the extraction of correct RVs was not possible due to unresolved but apparent components, the multiplicity could be revealed from the spectra. Fig.~\ref{fig:halpha} shows an example of an object, where RVs showed no variability, but spectral lines of both components were visible.

\begin{figure}
    \centering
    \includegraphics[width=0.93\linewidth]{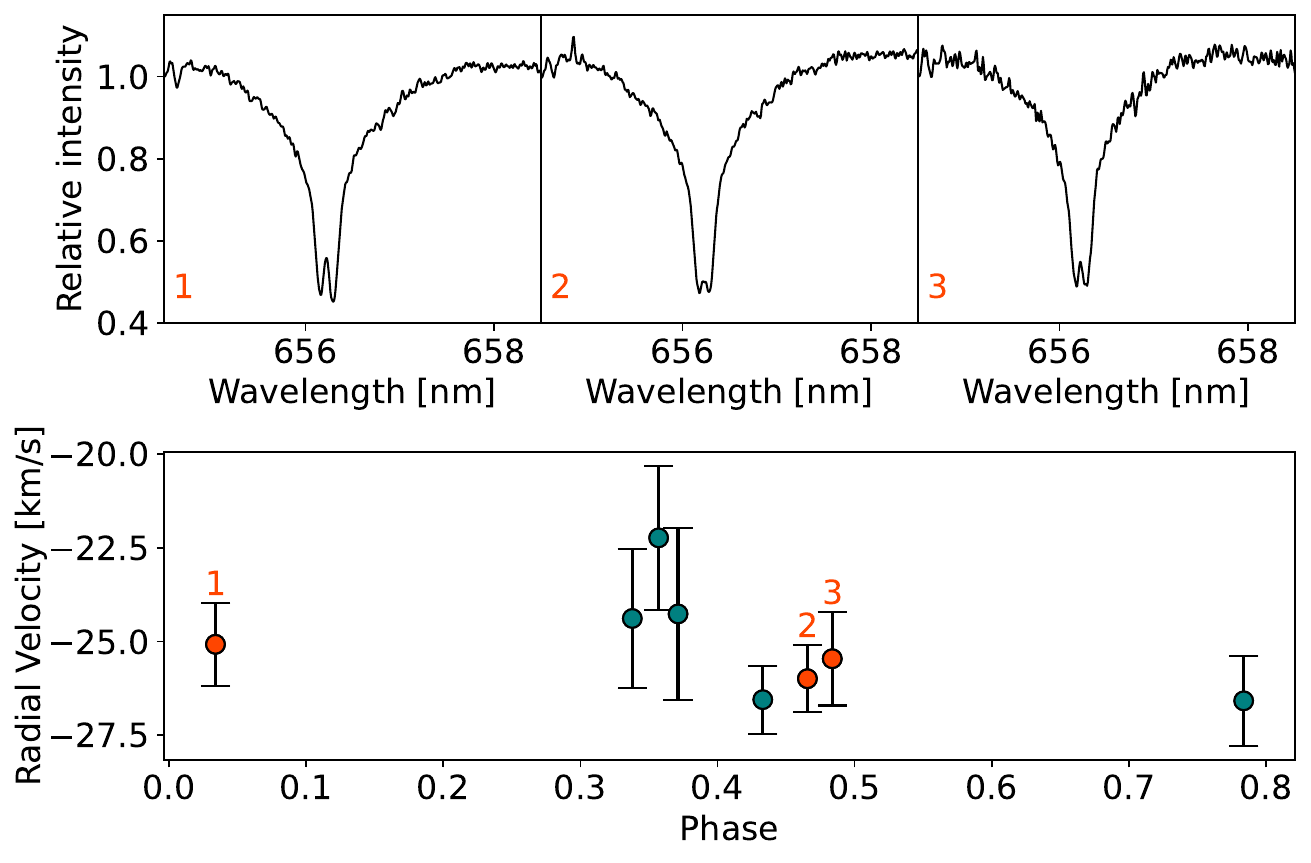}
    \caption{RV curve of TIC~174214184 measured with OES, where points highlighted with red show corresponding spectra zoomed on the $\mathrm{H\alpha}$ that shows lines of both components.}
    \label{fig:halpha}
\end{figure}

\subsection{Pulsating stars}
When a star pulsates, its outer layers periodically expand and contract in response to its pulsation mode, leading to a change in effective temperature. During compression, the material heats up and the effective temperature rises, increasing the brightness. During expansion, it cools and its brightness decreases.
This process introduces a phase difference of $\pi/4$ between the LC and RV curve.
In the sample of targets, pulsating variables, denoted Pu, were found by examining the phase shift of LC and RVs phased with $P$. These objects exhibited zero RV at maximum and minimum brightness, with peak (or trough) RV occurring midway between the flux extrema. An example of such a case from the sample can be seen in Fig.~\ref{fig:pu}.

\begin{figure}
    \centering
    \includegraphics[width=0.93\linewidth]{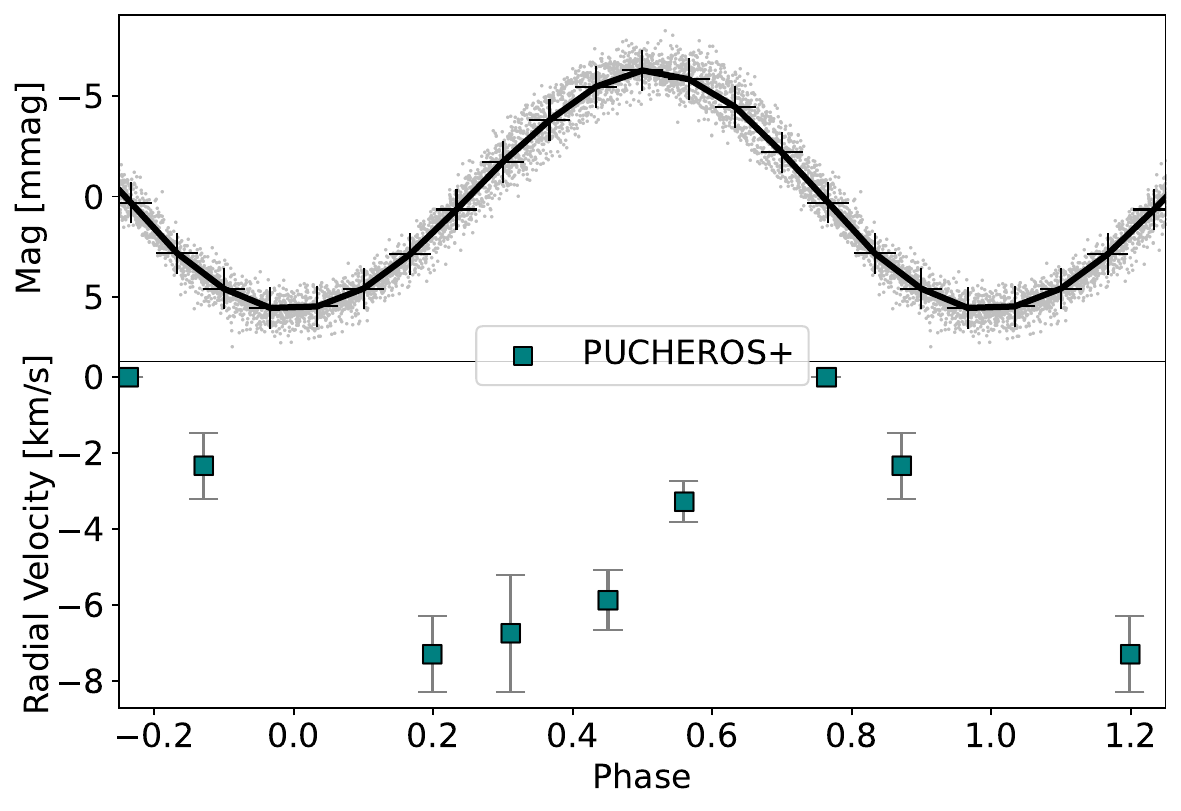}
    \caption{Example of LC (top) showing original (gray dots) and binned data (black crosses) and RV curve (bottom) for a pulsating star - TIC~205913291.}
    \label{fig:pu}
\end{figure}

\subsection{Candidates for spots}
Photometric spots in the atmospheres of hot stars redistribute radiation and lead to different LC shapes in different photometric filters, with variations on the order of millimagnitudes. Since the spots are confined to the rotating atmosphere, they do not induce a significant RV shift in the observed light. The RV variations caused by spots have very low amplitudes.
Spotted stars were identified based on the appearance of their RVs, supported by slight deformations of LCs. RV measurements were close to zero (see Fig.~\ref{fig:spot0}) with only slight variations in several cases (see Fig.~\ref{fig:spot1}).

\begin{figure}
    \centering
    \includegraphics[width=0.93\linewidth]{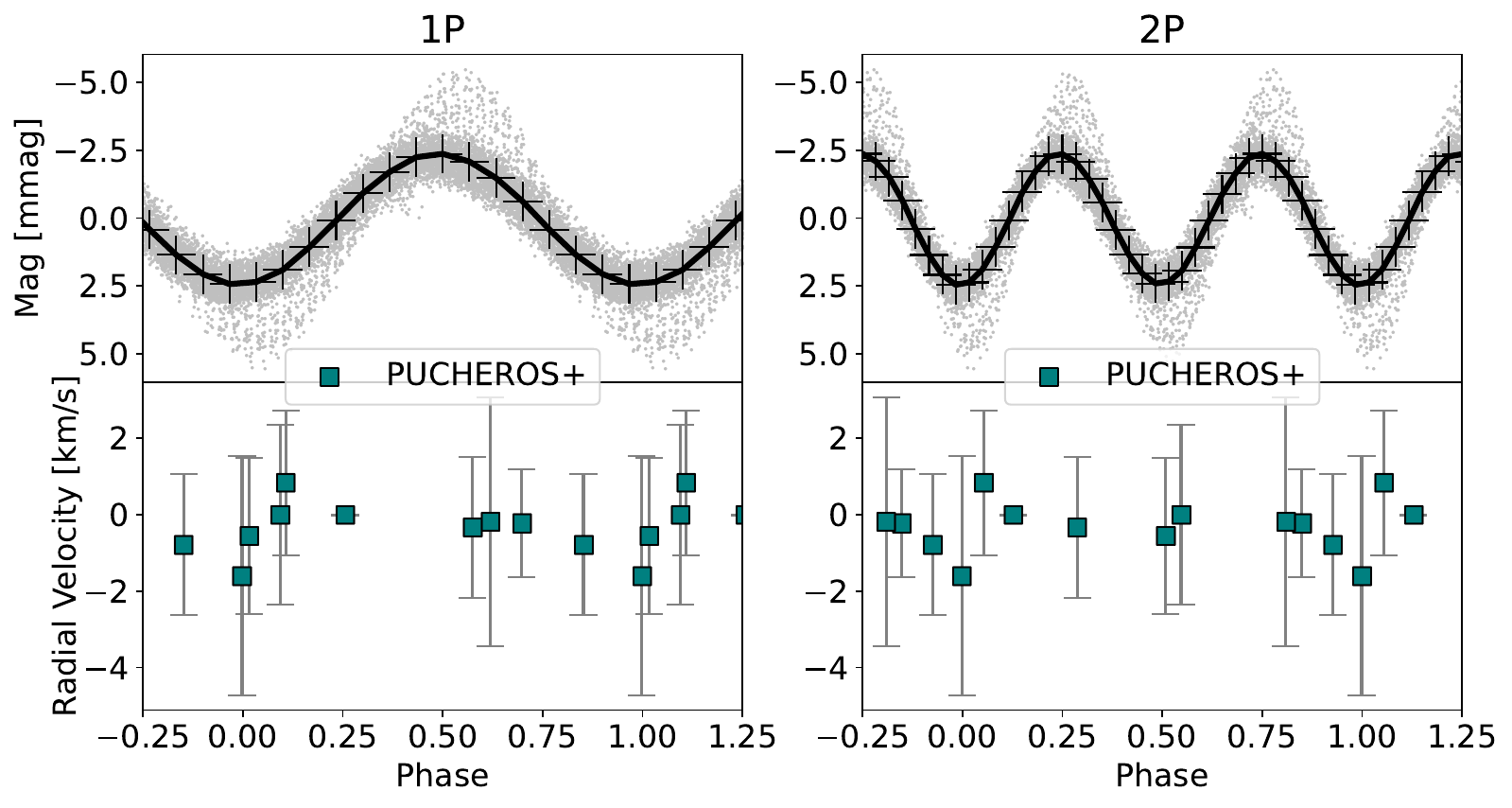}
    \caption{Example of LC (top panels) showing original (gray dots) and binned data (black crosses) and RV curve (bottom panels) for a candidate for spots with RVs close to zero - TIC~61449214.}
    \label{fig:spot0}
\end{figure}

\begin{figure}
    \centering
    \includegraphics[width=0.93\linewidth]{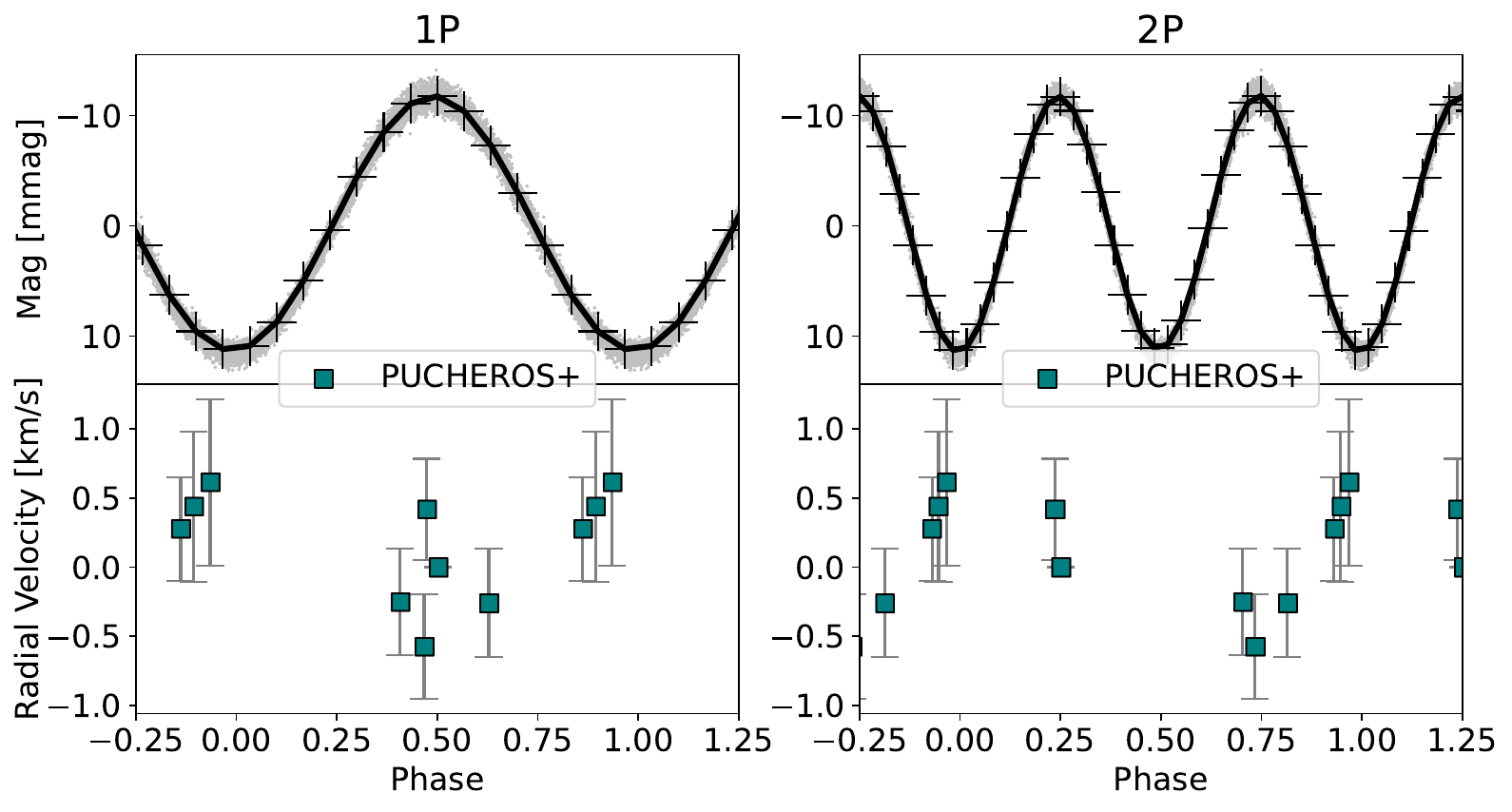}
    \caption{Example of LC (top panels) showing original (gray dots) and binned data (black crosses) and RV curve (bottom panels) for a candidate for spots with RVs close to zero - TIC~113150902.}
    \label{fig:spot1}
\end{figure}

Another way to produce photometric variability on the order of millimagnitudes with RVs close to zero is by observing a binary system with low inclination angles, nearly perpendicular to the orbital plane. These two cases cannot be distinguished with LCs and RVs alone, so they are marked as both "spots" and "binary". Their classification remains a matter of future work.

\subsection{Uncertain classification}
Even with the combination of photometric and spectroscopic data, the classification for some targets was unclear. These objects were attributed the "uncertain" tag for several reasons. First, the classification was not possible for objects with too few spectroscopic measurements (see Fig.~\ref{fig:uncertain1}). Similarly, the origin of the variations was unclear if the RVs showed variability when phased with both $P$ and $2P$. Fig.~\ref{fig:uncertain2} shows an example of such an object, where measurements in a specific phase could help determine the variability mechanism. Additionally, for some objects, the RV curve showed no periodic variations, but the scatter of the data was significant and could not be attributed to the noise (see Fig.~\ref{fig:uncertain3}). Lastly, the object was assigned an unclear classification if additional processing revealed it was a blend, and the variable signal originated from a nearby source.

\begin{figure}
    \centering
    \includegraphics[width=0.93\linewidth]{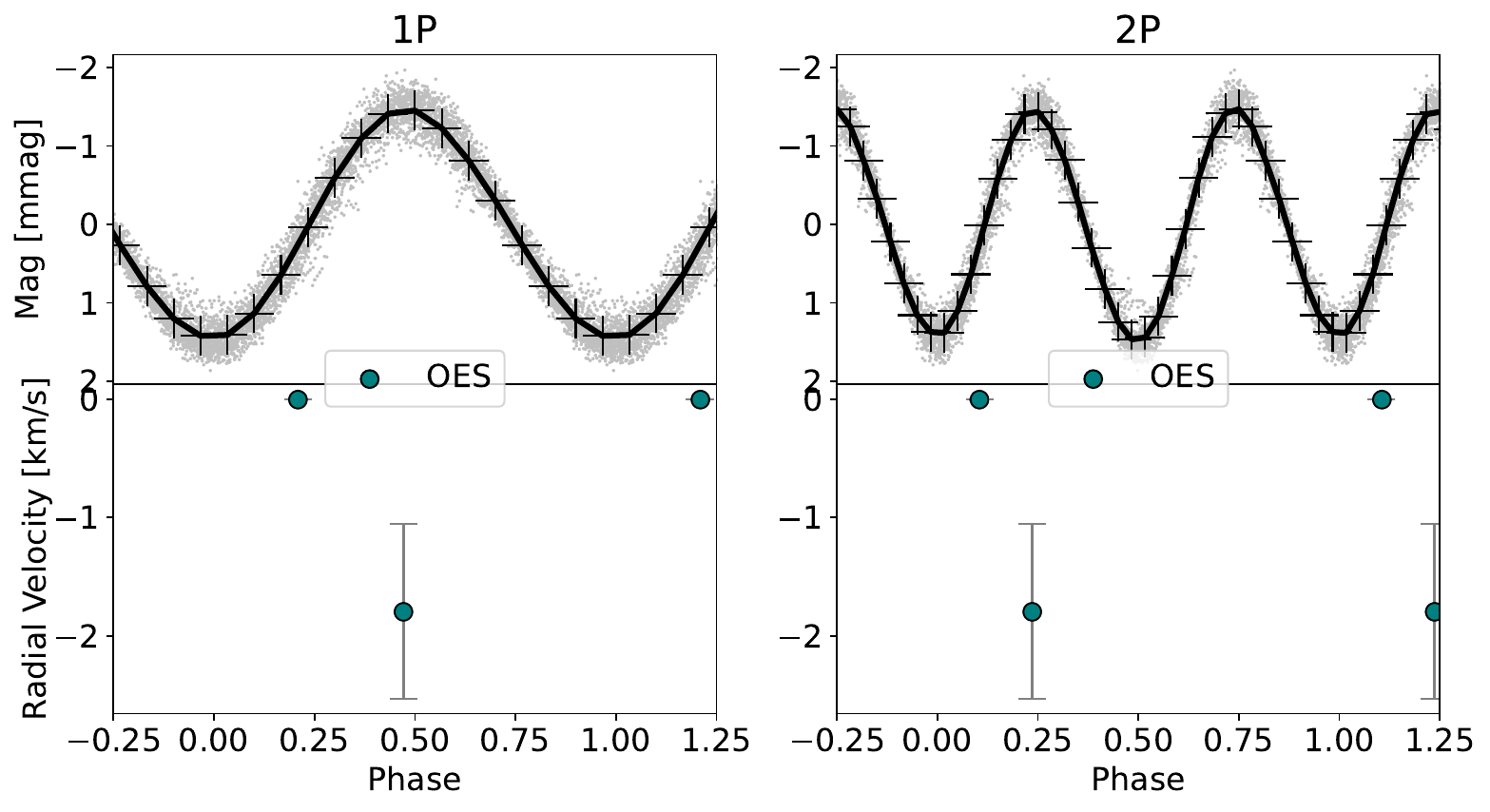}
    \caption{Example of LC (top panels) showing original (gray dots) and binned data (black crosses) and RV curve (bottom panels) for a star with uncertain classification with too few RV data points - TIC~88815918.}
    \label{fig:uncertain1}
\end{figure}

\begin{figure}
    \centering
    \includegraphics[width=0.93\linewidth]{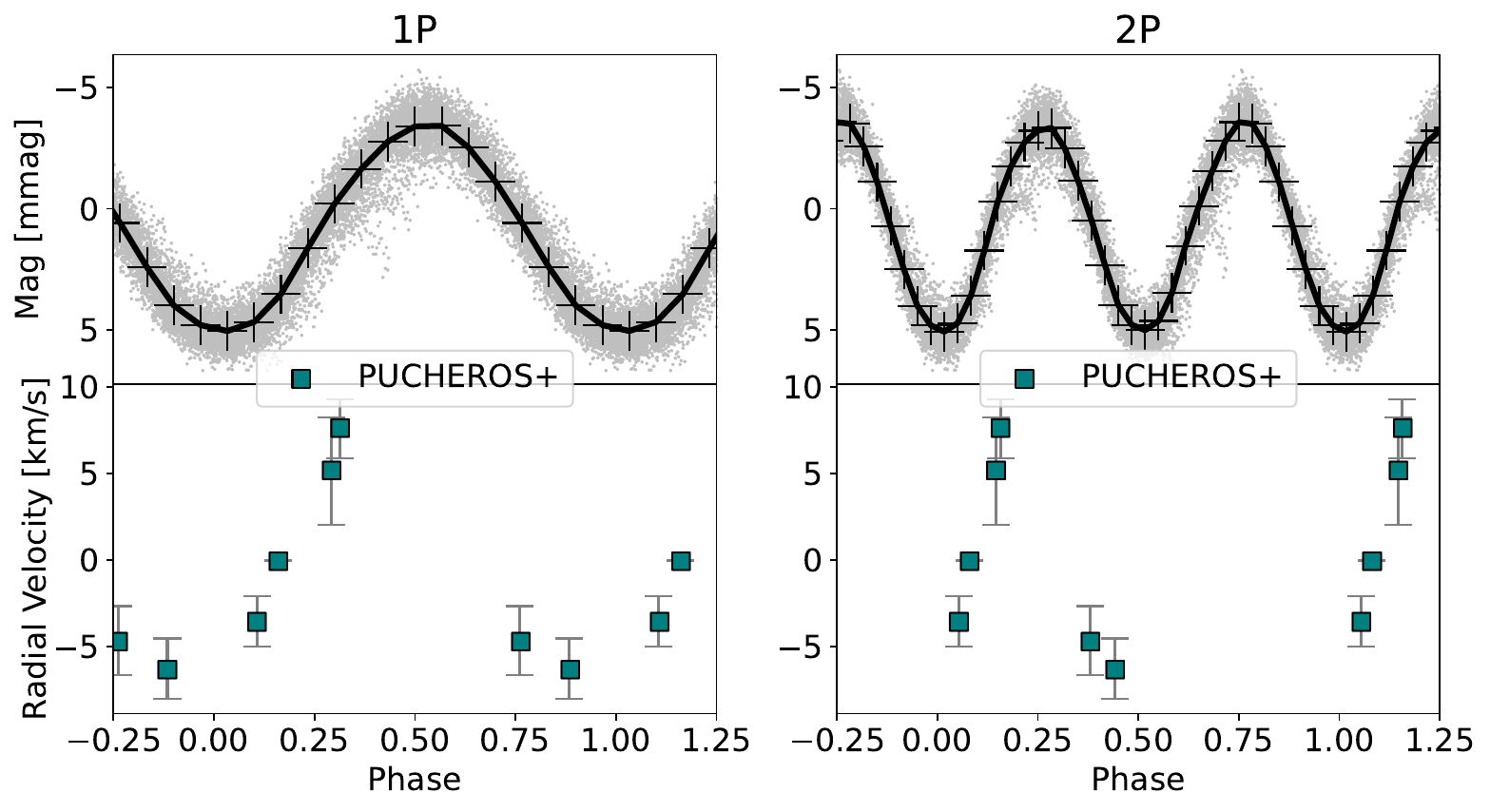}
    \caption{Example of LC (top panels) showing original (gray dots) and binned data (black crosses) and RV curve (bottom panels) for a star with uncertain classification, where we were unable to determine the variability mechanism due to the variability in both 1P and 2P - TIC~292207311.}
    \label{fig:uncertain2}
\end{figure}

\begin{figure}
    \centering
    \includegraphics[width=0.93\linewidth]{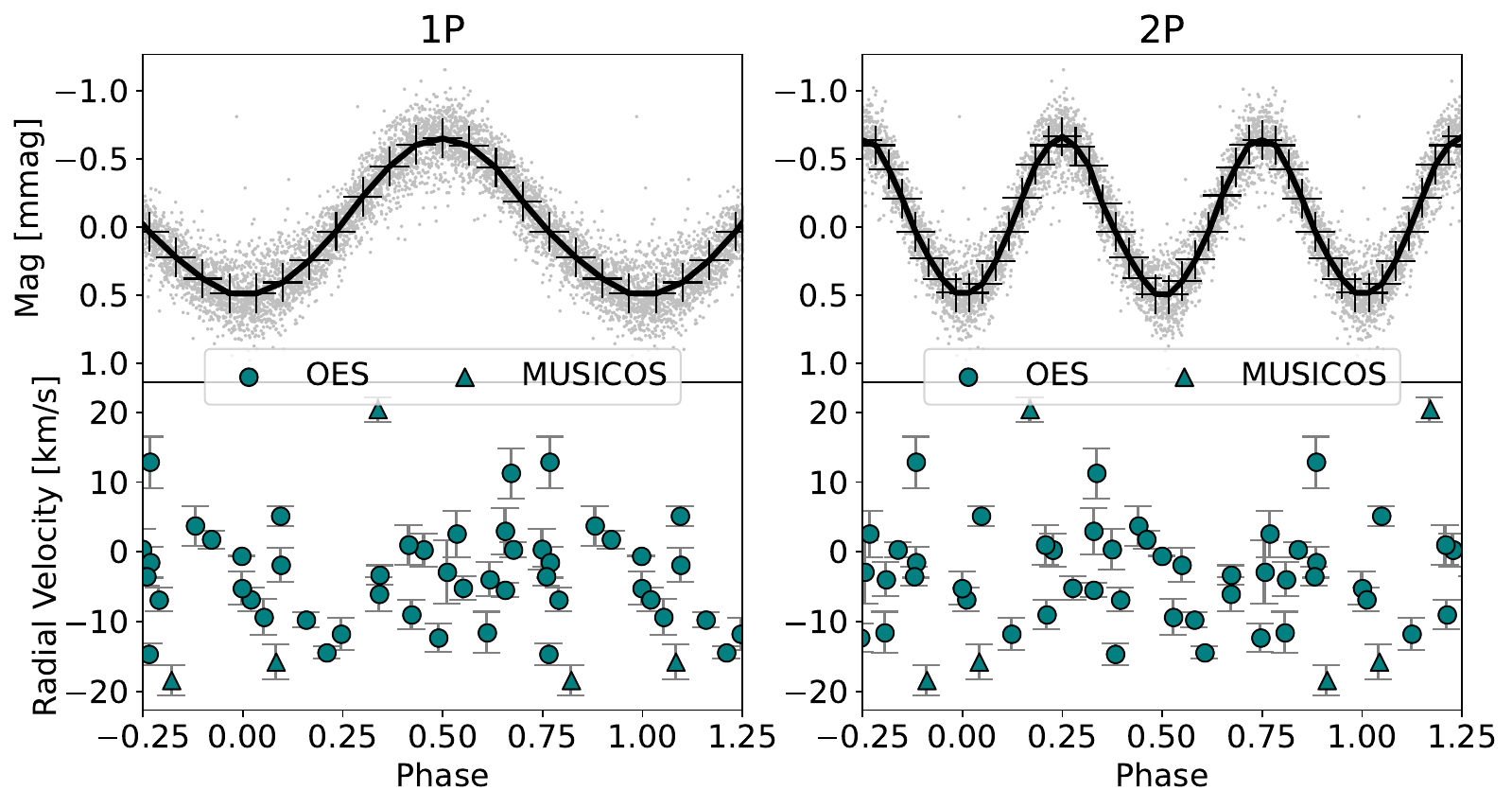}
    \caption{Example of LC (top panels) showing original (gray dots) and binned data (black crosses) and RV curve (bottom plots) for a star with uncertain classification with a large scatter of the data - TIC~16878120.}
    \label{fig:uncertain3}
\end{figure}

\section{Discussion}\label{Sect:Discussion}
Based on the classification criteria outlined in Sect.~\ref{Sect:Analysis}, we identified 18 binary stars, 1 pulsating star, 9 candidates for spotted stars, and 7 objects with uncertain classification (see table~\ref{Tab:Classification}). We conducted a detailed examination of sky background for each object using the in-built functions in the \textsc{Lightkurve} library, \textsc{TPF plotter} \citep{tpfplotter} and \textsc{TESS Localize} \citep{tesslocalize}. We searched for possible sources of contamination up to 6~mag fainter than the target, and calculated the origin of the variable signal using the data in individual pixels of TPFs. The origin of the variability was then associated with an object in the field of view (FoV). The vast majority of targets were identified as the origin of the variable signal, with only a single target identified as a blend (TIC\,444577764).

The final sample included 18 objects identified as binary systems. For eight systems, we confirmed the previous classification from Simbad (see Table~\ref{Tab:Classification}. 
In previous publications, TIC~220485766 was identified as either a pulsating or binary star \citep{gaiasb,gaiapu}, VSX lists this object as a pulsating star. Based on the spectroscopic data, we support the binary interpretation.
TIC~21673730 was previously identified as a pulsator \citep{216gd,vsx}. However, this classification was based only on photometric data from the SuperWASP survey of Am pulsators. Spectroscopic observations of this object revealed variability that can only be attributed to a binary system. 
The remaining 8 stars are newly identified SB1 or SB2 binary stars with no previous classification. Since the SB1/SB2 distinction depends on the quality of the observations, we assign these labels based solely on the available data and do not place significant emphasis on this division. Four of the sample stars identified as SB1 or SB2 are marked as metallic stars by \citep{Renson2009}, which is not surprising because about 70\,\% of metallic stars are found in binary systems \citep{Abt1961, Abt1985, Carquillat2007}. This chemical peculiarity might also contribute to light variations.

For all binary stars in the final sample, the observed variability might result from a combination of rotational modulation, reflection, Doppler beaming, and tidal interaction between the components. This complicates the retrieval of orbital parameters and makes detailed modelling beyond the scope of this work.

Based on the classification criteria, we identified only one pulsating object in the final sample. TIC~205913291 was marked in VSX \citep{vsx} as a variable star with unspecified variability type, identified in K2 observing campaign \citep{205k2}. This star is marked as magnetic CP star of Bp8 Si spectral type in \citep{Renson2009}. However, the amplitude of about 10\,km/s suggests that the radial-velocity variations are more likely caused by pulsations than by chemical spots.

Out of 35 objects with the spectroscopic follow-up, we identified 9 candidates for spotted stars. All objects are located on the southern hemisphere and have effective temperatures above 7000~K, consistent with the formation theory of stable photometric spots via diffusion and settling of different chemical elements \citep{Michaud1970,Michaud1976}. 

TIC~66497441 and TIC~184607315 exhibited long-term trends that could be attributed to a long-period orbital motion of a binary system around a common centre with a third body. After removing the long-term trend, the corrected RVs were inconclusive and the stars were classified as spots/binary+SB because we could not distinguish between rotational modulation due to spots and proximity effects in binary systems with low inclination. Confirming the presence of photometric spots on the surface of these stars would require detailed analysis of spectra, chemical abundances and multicolour photometry, which lies outside the scope of this work.

Uncertain objects comprise 20\,\% of the sample (see Table~\ref{Tab:Classification}). TIC~444577764 was identified as a blend during the background analysis mentioned above. This object was a part of a very crowded field, with a bright star blending into the aperture. VSX knows this object as rotationally variable star \citep{vsx}.
TIC~88815918 had only two spectroscopic observations with good SNR and many spectral lines. The object lacked the presence of a second companion in the spectral lines, even though the spectra were taken during quadrature. It is very likely that the object is not an SB2, however lack of additional data makes further classification impossible.

Similarly, TIC~310932102 shows RV variability when phased with $1P$. However, due to the scatter of the data, the lack of additional information makes the classification unclear. 
TIC~292207311 and TIC~448876509, known as variable stars in VSX \citep{vsx}, showed variability when phased with both $P$ and $2P$ but lacked phase coverage between 0.5 and 1.0 which made the determination of variation mechanism impossible. 
TIC~16878120 and TIC~218160121 both exhibited a large scatter of the data (the standard deviation was 8.2\,km/s and 6.0\,km/s, respectively) that could not be attributed to noise. TIC~16878120 additionally exhibited a long-term trend that could not be removed due to poor data coverage. This trend could be explained by instrumental effects or changes in observing conditions; however, given the consistency of the RVs obtained from OES and MUSICOS under different conditions and instrumental setups, we are confident in the reliability of the measurements. The observed scatter in RVs could be attributed to intrinsic long-term variability. However, the dataset does not provide sufficient coverage to support this hypothesis, making fitting and correction of RVs impossible. In case of TIC~218160121, the scatter in RVs might be due to chemical peculiarity \citep{Renson2009}.

\section{Conclusions}\label{Sect:Conclusions}

We identified a sample of 108 F to O-type main-sequence stars with sinusoidal brightness variations on the level of millimagnitudes that can be explained by different variability mechanisms -- as binarity, pulsations or rotational variability due to photometric spots. This work highlights the complexity of stellar classification and the need for detailed individual approach.

We employed a careful, multistep selection process involving automated scripts and manual inspection to select only stars with clean monotonic sinusoidal variations. These stars were spectroscopically observed by using three spectrographs; OES and MUSICOS in the northern hemisphere and PUCHEROS+ in the southern hemisphere. We obtained spectra for 35 objects in the sample and used custom \textsc{Iraf} scripts to extract RVs of these targets. The sky background of each object was carefully analysed with Python routines, and the contamination from nearby stars was considered.

The analysed targets were divided into four categories based on the appearance of their phase-folded LCs with respect to phase-folded RVs: binary systems, pulsating stars, spotted star candidates and objects with uncertain classification (Table~\ref{Tab:Classification}). These categories reflect the characteristics of the chosen dataset, mainly the small amplitude of photometric variations and variability linked to processes in the radiative envelopes of hot stars. Binary stars comprised more than 50~\% of the analysed targets. This is not surprising, as the multiplicity fraction, depending on stellar environment, increases with stellar mass \citep{chini}. \citet{duchene} report the multiplicity fraction of $70\pm9$\,\% for O-type stars, $45\pm5$\,\% for B-type stars and 30-40~\% for stars between spectral types B5 and F.

Candidates for spotted stars represent one fourth of our sample stars. However, this number could change with the proper identification of rotational variability through detailed analysis of chemical abundances and multicolour photometry. In literature, chemically peculiar stars with stable photometric spots in radiative atmospheres make up about 10~\% of B to F-type stars \citep{hummerich}. We spotted only one pulsating star in our sample. The rest of the sample (about 20~\%) was assigned an uncertain classification. With additional spectroscopic measurements, the variation mechanism of some of these objects could be resolved. 

Our study revealed that from randomly selected stars with sinusoidal photometric variations, about half of them are binary stars. It is worth noting that we did not find any overlap between our sample and catalogues of chemically peculiar stars from LAMOST \citep{Ghazaryan2018,Qin2019}, and no overlap with magnetic stars \citep{Rustem2023}. Eight stars were found in the catalogue by \citet{Renson2009}.  TIC\,205913291 and TIC\,226037840 are labelled as Bp9 Si and Bp8 Si, respectively, which points towards the magnetic chemically peculiar nature of these stars.

Nineteen stars from our sample have a remark about their variability in literature. However, our classification is in at least partial agreement with literature in only nine of these stars. We revealed eight new binary systems, one new pulsating star and nine candidates for spotted stars. However, these spot candidates might include binary systems with low inclination angles. These results highlight the discrepancies in catalogue values and the necessity of careful analysis and follow-up spectroscopic measurements, especially for objects where photometric data alone are insufficient for determining the variability mechanism.

\begin{acknowledgements}
      We would like to thank the observers for their work. We acknowledge a bilateral mobility project between Slovak Academy of Sciences and Czech Academy of Sciences SAV-25-10. MV and TP acknowledge support from the Slovak Research and Development Agency – contract No. APVV-20-0148, and the VEGA grant of the Slovak Academy of Sciences No. 2/0031/22. MS acknowledges the support by Inter-transfer grant no LTT-20015. This paper includes data collected with the TESS mission and with the Perek telescope at the Astronomical Institute of the Czech Academy of Sciences in Ond\v{r}ejov. Funding for the TESS mission is provided by the NASA Explorer Program. The TESS data were obtained from the MAST data archive at the Space Telescope Science Institute (STScI). This research made use of NASA’s Astrophysics Data System Bibliographic Services, and of the SIMBAD database, operated at CDS, Strasbourg, France. 
\end{acknowledgements}

\bibliographystyle{aa}
\bibliography{references}

\begin{appendix} \label{Sect:Appendix}
\onecolumn
\section{Classification table}
\begin{table}[h!]
\centering
\caption{Classification of our sample of stars. The column 'Type' gives classification in this work, column 'CP' gives the spectral type from \citet{Renson2009}, column 'VSX' gives the classification in \citet{vsx}. The other columns are self-explanatory.}
\label{Tab:Classification}
\begin{tabular}{ccccccccl}
\hline \hline
TIC &
\begin{tabular}{c} RA \\ (deg) \end{tabular} &
\begin{tabular}{c} DEC \\ (deg) \end{tabular} &
\begin{tabular}{c} $f$ \\ (c/d) \end{tabular} &
\begin{tabular}{c} $P$ \\ (d) \end{tabular} &
\begin{tabular}{c} $T_\mathrm{eff}$ \\ (K) \end{tabular} &
Type &
CP &
VSX type \\
\hline
5638336	&	32.327	&	28.394	&	0.469	&	2.132	&	9212	&	SB1	&	\ldots & \ldots\\
12321432	&	32.835	&	64.146	&	0.370	&	2.703	&	13491	&	SB2	& \ldots &\ldots \\
14400891	&	127.769	&	54.081	&	0.435	&	2.299	&	6790	&	SB2	&\ldots &\ldots\\
16878120	&	232.692	&	34.466	&	3.187	&	0.314	&	9677	&	uncertain	&\ldots&\ldots\\
21673730	&	255.722	&	31.690	&	0.623	&	1.605	&	7879	&	SB1	&	A2-	&	GDOR	\\
31623275	&	289.076	&	-34.446	&	1.250	&	0.800	&	10850	&	SB1	&\ldots&	ACV|roAm|roAp|SXARI	\\
61449214	&	228.484	&	-26.051	&	1.549	&	0.646	&	7697	&	spots/binary	&\ldots &\ldots\\
66497441	&	356.050	&	-18.277	&	2.612	&	0.383	&	11505	&	spots/binary+SB	&\ldots&\ldots\\
84756974	&	249.991	&	-39.789	&	3.567	&	0.280	&	9265	&	spots/binary	&\ldots&\ldots\\
88815918	&	27.046	&	16.956	&	0.836	&	1.196	&	10471	&	uncertain	&\ldots&\ldots\\
113150902	&	277.024	&	-38.357	&	0.990	&	1.010	&	11200	&	spots/binary	&\ldots&	VAR	\\
135081803	&	184.888	&	-40.839	&	0.750	&	1.333	&	7861	&	SB2	&\ldots &\ldots\\
137800207	&	344.066	&	-23.849	&	1.661	&	0.602	&	7023	&	spots/binary	&\ldots&\ldots\\
160644410	&	223.727	&	-36.430	&	0.671	&	1.490	&	9425	&	spots/binary	&\ldots&\ldots\\
174214184	&	234.737	&	-41.098	&	0.937	&	1.067	&	6914	&	SB2	&	A1-F2	&\ldots\\
184607315	&	31.603	&	38.900	&	0.814	&	1.229	&	9390	&	spots/binary+SB	&\ldots&\ldots\\
205913291	&	250.398	&	-28.586	&	0.783	&	1.277	&	10853	&	Pu	&	B9 Si	&	VAR	\\
212031970	&	248.419	&	-31.289	&	1.887	&	0.530	&	9688	&	SB2	&\ldots&	VAR	\\
218160121	&	261.994	&	-44.721	&	4.568	&	0.219	&	7514	&	uncertain	&	A1-	&\ldots\\
220485766	&	343.501	&	-54.727	&	1.254	&	0.797	&	8715	&	SB1	&\ldots&	DSCT|GDOR|SXPHE	\\
226037840	&	247.367	&	-46.261	&	2.013	&	0.497	&	10393	&	spots/binary	&	B8 Si	&	MISC	\\
257456854	&	257.821	&	24.252	&	0.886	&	1.129	&	6750	&	SB2	&\ldots&\ldots\\
279821618	&	239.879	&	-40.865	&	2.181	&	0.459	&	8574	&	spots/binary	&\ldots&\ldots\\
292207311	&	230.933	&	-27.952	&	1.264	&	0.791	&	7317	&	uncertain	&\ldots&\ldots\\
293069615	&	254.137	&	-66.109	&	0.795	&	1.258	&	8609	&	SB2	&\ldots&\ldots\\
302581695	&	133.482	&	35.538	&	0.986	&	1.014	&	9101	&	SB1	&\ldots&	ELL	\\
302666414	&	163.005	&	55.355	&	0.623	&	1.605	&	7419	&	SB1	&	A8-F3	&\ldots\\
310932102	&	262.096	&	29.838	&	0.767	&	1.304	&	9955	&	uncertain	&\ldots&\ldots\\
320692159	&	271.163	&	27.115	&	1.583	&	0.632	&	7646	&	SB2	&\ldots&	VAR	\\
342829903	&	205.359	&	-67.882	&	0.633	&	1.580	&	6666	&	SB2	&\ldots&\ldots\\
347699402	&	263.048	&	21.264	&	0.304	&	3.289	&	8520	&	SB1	&\ldots&\ldots\\
351532879	&	309.160	&	-63.121	&	0.486	&	2.058	&	6658	&	SB2	&\ldots&\ldots\\
419610625	&	320.367	&	-66.667	&	0.718	&	1.393	&	7832	&	SB1	&	A3-F0	&\ldots\\
444577764	&	267.783	&	-35.337	&	1.612	&	0.620	&	11608	&	uncertain	&\ldots&	ACV|roAm|roAp|SXARI	\\
448876509	&	214.291	&	-68.407	&	0.347	&	2.882	&	11798	&	uncertain	&\ldots&	VAR	\\ \hline \hline
\end{tabular}
\end{table}

\pagebreak
\section{Measured relative radial velocities}
\begin{table}[ht!]
\centering
\scriptsize
\caption{Relative radial velocities for targets in our sample of stars}
\label{Tab:obs01}
\begin{tabular}{ccc ccc ccc}
\hline \hline
HJD & 
\begin{tabular}{c} RV \\ ($\mathrm{km\,s^{-1}}$) \end{tabular} &
\begin{tabular}{c} err \\ ($\mathrm{km\,s^{-1}}$) \end{tabular} &
HJD &
\begin{tabular}{c} RV \\ ($\mathrm{km\,s^{-1}}$) \end{tabular} &
\begin{tabular}{c} err \\ ($\mathrm{km\,s^{-1}}$) \end{tabular} &
HJD &
\begin{tabular}{c} RV \\ ($\mathrm{km\,s^{-1}}$) \end{tabular} &
\begin{tabular}{c} err \\ ($\mathrm{km\,s^{-1}}$) \end{tabular} \\
\hline
\multicolumn{3}{c}{TIC 5638336 (OES)} & \multicolumn{3}{c}{TIC 16878120 (OES)} & \multicolumn{3}{c}{TIC 84756974 (PUCHEROS+)} \\
2460559.5210 & 0.00 & 0.00 & 2460774.4907 & 0.31 & 1.10 & 2460528.6413 & -0.59 & 6.00 \\
2460576.6100 & -0.87 & 0.43 & 2460776.5042 & 5.10 & 1.40 & 2460544.5228 & -3.29 & 11.00 \\
2460663.3410 & 18.59 & 0.22 & 2460794.5276 & 2.55 & 3.30 & 2460545.6120 & 3.85 & 7.90 \\
2460672.3449 & 34.48 & 0.31 & 2460795.5069 & 2.93 & 3.30 & 2460559.5409 & -5.36 & 5.50 \\
2460673.3094 & 41.16 & 0.30 & 2460796.4835 & -1.49 & 2.20 & 2460567.5269 & -5.14 & 8.60 \\
2460674.3121 & 11.54 & 0.46 & 2460797.4726 & 1.76 & 1.30 & \multicolumn{3}{c}{TIC 88815918 (OES)} \\
\multicolumn{3}{c}{TIC 12321432 (OES)} & 2460805.5475 & -5.45 & 1.00 & 2460559.5065 & 0.00 & 0.00 \\
2460559.5636 & -6.53 & 3.50 & 2460809.4505 & -1.87 & 2.50 & 2460576.5639 & -1.79 & 0.74 \\
2460672.3793 & 74.86 & 5.30 & 2460826.4716 & -3.29 & 1.40 & \multicolumn{3}{c}{TIC 113150902 (PUCHEROS+)} \\
2460673.3545 & -9.25 & 2.70 & 2460828.4851 & -3.50 & 1.30 & 2460495.8025 & 0.00 & 0.00 \\
2460674.3597 & 73.56 & 5.50 & 2460836.4545 & -9.72 & 1.10 & 2460499.8156 & 0.42 & 0.37 \\
2460683.3540 & 67.54 & 6.40 & 2460840.4902 & -6.82 & 1.70 & 2460500.7593 & -0.25 & 0.39 \\
\multicolumn{3}{c}{TIC 14400891 (OES)} & 2460844.3815 & -8.98 & 2.10 & 2460535.5768 & 0.28 & 0.37 \\
2460428.3782 & -80.88 & 0.61 & 2460846.3796 & -6.86 & 1.70 & 2460543.7352 & 0.61 & 0.60 \\
2460428.3782 & 93.27 & 0.64 & 2460847.3862 & -5.21 & 2.40 & 2460545.7148 & 0.44 & 0.54 \\
2460428.3782 & 44.18 & 1.60 & 2460850.4046 & -3.94 & 2.40 & 2460555.5517 & -0.26 & 0.39 \\
2460429.4306 & -14.65 & 0.58 & 2460855.3846 & -12.29 & 2.00 & 2460573.5788 & -0.57 & 0.38 \\
2460429.4306 & 32.13 & 0.48 & 2460856.3453 & -5.17 & 1.70 & \multicolumn{3}{c}{TIC 135081803 (PUCHEROS+)} \\
2460430.3853 & -72.07 & 0.55 & \multicolumn{3}{c}{TIC 16878120 (MUSICOS)} & 2460516.4642 & -106.71 & 2.00 \\
2460430.3853 & 82.25 & 0.43 & 2460712.5716 & 20.39 & 1.80 & 2460516.4642 & 72.70 & 2.70 \\
2460430.3853 & 41.29 & 1.30 & 2460859.3360 & -15.73 & 2.50 & 2460518.5051 & -87.05 & 2.00 \\
2460431.3718 & -52.44 & 0.69 & 2460879.3352 & -18.37 & 2.20 & 2460518.5051 & 54.79 & 2.90 \\
2460431.3718 & 63.30 & 0.58 & \multicolumn{3}{c}{TIC 21673730 (OES)} & \multicolumn{3}{c}{TIC 137800207 (PUCHEROS+)} \\
2460431.3718 & 38.73 & 1.50 & 2460428.4812 & 56.57 & 0.68 & 2460494.9256 & 2.80 & 3.50 \\
2460433.3597 & -69.98 & 0.44 & 2460430.4977 & -51.01 & 0.56 & 2460495.8530 & 4.13 & 5.20 \\
2460433.3597 & 87.81 & 0.52 & 2460431.4725 & 58.49 & 0.40 & 2460499.9003 & 6.07 & 2.40 \\
2460433.3597 & 32.85 & 1.30 & 2460433.4597 & -79.26 & 0.59 & 2460500.9409 & 5.12 & 2.70 \\
2460435.3190 & -80.25 & 0.57 & 2460435.4204 & 0.05 & 0.04 & 2460510.7198 & 3.46 & 2.70 \\
2460435.3190 & 95.05 & 0.45 & 2460511.3746 & 37.66 & 0.87 & 2460518.6863 & 2.74 & 3.20 \\
2460435.3190 & 30.78 & 1.40 & 2460512.3762 & 7.19 & 0.74 & 2460544.6784 & -0.24 & 3.70 \\
2460436.3255 & -9.37 & 0.77 & 2460520.3333 & -53.80 & 0.51 & 2460545.7895 & -0.92 & 3.80 \\
2460436.3255 & 28.02 & 0.41 & 2460523.3454 & -75.34 & 0.63 & 2460559.8369 & 3.02 & 3.90 \\
2460709.4034 & -57.62 & 0.51 & 2460526.3528 & -90.85 & 0.78 & 2460565.7616 & 0.76 & 4.10 \\
2460709.4034 & 71.92 & 0.64 & 2460526.3617 & -90.13 & 0.60 & 2460568.7572 & 0.00 & 0.04 \\
2460709.4034 & 25.81 & 1.10 & 2460559.2686 & 9.19 & 0.76 & \multicolumn{3}{c}{TIC 160644410 (PUCHEROS+)} \\
2460711.2552 & -79.62 & 0.54 & \multicolumn{3}{c}{TIC 21673730 (MUSICOS)} & 2460516.5284 & -1.13 & 1.10 \\
2460711.2552 & 93.41 & 0.45 & 2460830.4102 & -47.81 & 1.90 & 2460518.5704 & -1.05 & 0.46 \\
2460711.2552 & 22.54 & 0.70 & 2460833.4144 & -17.65 & 1.90 & 2460520.4925 & -0.38 & 1.10 \\
2460715.3399 & -65.63 & 0.52 & 2460836.4197 & 11.43 & 2.00 & 2460569.5185 & 0.01 & 0.02 \\
2460715.3399 & 85.20 & 0.50 & 2460838.4285 & 10.35 & 2.30 & \multicolumn{3}{c}{TIC 174214184 (PUCHEROS+)} \\
2460715.3399 & 10.80 & 1.30 & 2460848.4137 & 49.56 & 2.00 & 2460517.5187 & -24.26 & 2.30 \\
2460716.2484 & -57.77 & 0.58 & 2460849.3998 & -10.80 & 1.50 & 2460519.5826 & -24.39 & 1.90 \\
2460716.2484 & 79.01 & 0.48 & \multicolumn{3}{c}{TIC 31623275 (PUCHEROS+)} & 2460530.5308 & -25.99 & 0.90 \\
2460716.2484 & 9.30 & 1.40 & 2460495.8278 & 0.00 & 0.00 & 2460534.5693 & -22.23 & 1.90 \\
2460716.2767 & -54.21 & 0.60 & 2460499.8362 & -27.98 & 1.70 & 2460545.5152 & -25.46 & 1.30 \\
2460716.2767 & 75.53 & 0.51 & 2460532.6537 & -0.38 & 1.30 & 2460547.5421 & -26.55 & 0.91 \\
2460716.2767 & 10.53 & 1.40 & 2460536.5226 & -28.80 & 1.50 & 2460559.5006 & -25.08 & 1.10 \\
2460722.2462 & -59.70 & 0.37 & 2460543.7584 & -4.60 & 1.70 & 2460567.5070 & -26.58 & 1.20 \\
2460722.2462 & 92.21 & 0.52 & 2460545.7314 & -20.64 & 2.40 & \multicolumn{3}{c}{TIC 184607315 (OES)} \\
2460722.2462 & -4.34 & 1.10 & 2460573.5987 & -25.18 & 1.20 & 2460559.4828 & 0.00 & 0.00 \\
2460722.2746 & -60.84 & 0.41 & 2460574.6109 & -12.44 & 1.70 & 2460576.5119 & -5.00 & 0.37 \\
2460722.2746 & 93.45 & 0.53 & \multicolumn{3}{c}{TIC 61449214 (PUCHEROS+)} & 2460583.4688 & -7.62 & 0.44 \\
2460722.2746 & -3.11 & 1.20 & 2460516.5476 & -0.78 & 1.90 & 2460663.2956 & -42.26 & 0.39 \\
2460726.2386 & 1.62 & 0.48 & 2460518.5898 & -0.55 & 2.10 & 2460672.2996 & -45.16 & 0.40 \\
2460726.2386 & 37.86 & 0.65 & 2460520.5145 & -1.60 & 3.10 & 2460673.3315 & -45.55 & 0.43 \\
2460726.2386 & -7.23 & 1.40 & 2460531.5495 & -0.01 & 2.40 & 2460674.3367 & -45.69 & 0.42 \\
\multicolumn{3}{c}{TIC 14400891 (MUSICOS)} & 2460565.5034 & -0.22 & 1.40 & 2460683.3312 & -48.13 & 0.50 \\
2460712.3107 & 6.83 & 0.13 & 2460572.5242 & -0.33 & 1.90 & \multicolumn{3}{c}{TIC 205913291 (PUCHEROS+)} \\
2460717.4384 & -44.28 & 0.40 & 2460573.5141 & 0.84 & 1.90 & 2460514.5770 & -2.33 & 0.87 \\
2460717.4384 & 71.81 & 0.52 & 2460574.4891 & -0.18 & 3.30 & 2460520.5609 & -3.28 & 0.53 \\
2460717.4384 & 6.65 & 1.20 & 2460577.4826 & -0.01 & 0.01 & 2460530.6331 & -5.87 & 0.79 \\
2460731.5141 & -52.44 & 0.50 & \multicolumn{3}{c}{TIC 66497441 (PUCHEROS+)} & 2460544.4944 & -6.74 & 1.50 \\
2460731.5141 & 104.70 & 0.90 & 2460495.9073 & 30.15 & 1.80 & 2460545.6276 & -7.30 & 1.00 \\
\multicolumn{3}{c}{TIC 16878120 (OES)} & 2460499.9357 & 29.85 & 0.85 & 2460556.5585 & -0.01 & 0.00 \\
2460428.4665 & 3.70 & 2.90 & 2460499.9416 & 30.05 & 1.20 & \multicolumn{3}{c}{TIC 212031970 (PUCHEROS+)} \\
2460429.5872 & 0.28 & 2.40 & 2460518.7049 & 25.16 & 1.20 & 2460520.5389 & -81.98 & 3.20 \\
2460431.4580 & 1.00 & 2.90 & 2460532.6977 & 19.87 & 1.30 & 2460520.5389 & 76.28 & 3.70 \\
2460433.4453 & 0.36 & 2.90 & 2460533.7867 & 18.47 & 1.20 & 2460530.6136 & -68.60 & 2.80 \\
2460435.4059 & -0.58 & 0.02 & 2460543.8289 & 14.39 & 4.30 & 2460530.6136 & 72.01 & 4.60 \\
2460511.3557 & -9.34 & 2.60 & 2460545.8049 & 14.15 & 0.84 & 2460544.4748 & -19.78 & 2.60 \\
2460512.3577 & -11.74 & 2.30 & 2460558.8243 & 8.09 & 1.20 & 2460544.4748 & 22.41 & 4.30 \\
2460520.3164 & -11.55 & 3.00 & 2460559.8609 & 6.88 & 1.10 & 2460545.5838 & -61.35 & 36.00 \\
2460523.3282 & -14.42 & 0.90 & 2460569.7355 & 2.11 & 1.40 & 2460545.5838 & 9.00 & 4.70 \\
2460526.3264 & -14.63 & 1.60 & 2460573.6427 & 1.39 & 1.00 & 2460556.5390 & -98.69 & 3.10 \\
2460738.6690 & -2.87 & 4.50 & 2460575.6316 & -0.03 & 0.31 & 2460556.5390 & 93.53 & 2.90 \\
2460740.6321 & 12.83 & 3.70 & \multicolumn{3}{c}{TIC 84756974 (PUCHEROS+)} & 2460566.5810 & -119.34 & 3.80 \\
2460743.6360 & -6.03 & 2.40 & 2460514.5570 & 0.00 & 0.00 & 2460566.5810 & 112.84 & 5.00 \\
2460746.5634 & 11.24 & 3.60 & 2460518.6109 & -0.60 & 5.10 &  &  & \\\hline \hline
\end{tabular}
\end{table}

\setcounter{table}{0}
\begin{table}[]
\centering
\scriptsize
\caption{continued}
\begin{tabular}{ccccccccc}
\hline \hline
HJD & 
\begin{tabular}{c} RV \\ ($\mathrm{km\,s^{-1}}$) \end{tabular} &
\begin{tabular}{c} err \\ ($\mathrm{km\,s^{-1}}$) \end{tabular} &
HJD &
\begin{tabular}{c} RV \\ ($\mathrm{km\,s^{-1}}$) \end{tabular} &
\begin{tabular}{c} err \\ ($\mathrm{km\,s^{-1}}$) \end{tabular} &
HJD &
\begin{tabular}{c} RV \\ ($\mathrm{km\,s^{-1}}$) \end{tabular} &
\begin{tabular}{c} err \\ ($\mathrm{km\,s^{-1}}$) \end{tabular} \\
\hline
\multicolumn{3}{c}{TIC 218160121 (PUCHEROS+)} & \multicolumn{3}{c}{TIC 293069615 (PUCHEROS+)} & \multicolumn{3}{c}{TIC 347699402 (MUSICOS)} \\ 
2460439.8756 & -77.89 & 12.00 & 2460519.6869 & 79.76 & 1.10 & 2460879.3642 & -0.65 & 0.87 \\
2460440.8527 & -82.57 & 15.00 & 2460520.5922 & -68.54 & 1.80 & \multicolumn{3}{c}{TIC 351532879 (PUCHEROS+)} \\
2460499.7717 & -40.08 & 14.00 & 2460520.5922 & 62.77 & 1.80 & 2460499.8550 & -78.48 & 2.60 \\
2460535.5536 & -47.99 & 15.00 & 2460530.5923 & -55.96 & 1.20 & 2460499.8550 & 90.35 & 3.20 \\
2460545.6815 & -48.99 & 15.00 & 2460530.5923 & 51.90 & 1.30 & 2460500.8083 & -0.31 & 1.50 \\
2460555.5300 & -48.12 & 12.00 & 2460535.5288 & -45.83 & 1.10 & 2460510.6697 & -81.39 & 4.10 \\
2460573.5580 & -47.70 & 0.54 & 2460535.5288 & 38.60 & 1.30 & 2460510.6697 & 80.36 & 2.10 \\
\multicolumn{3}{c}{TIC 220485766 (PUCHEROS+)} & 2460545.6450 & -52.79 & 1.00 & 2460513.8270 & -40.34 & 6.20 \\
2460495.8828 & -28.20 & 0.97 & 2460545.6450 & 48.05 & 1.10 & 2460513.8270 & 42.89 & 2.80 \\
2460499.8788 & 10.26 & 1.20 & 2460566.6023 & -36.88 & 1.00 & 2460518.6550 & -95.84 & 4.00 \\
2460500.8955 & -21.63 & 2.10 & 2460566.6023 & 31.15 & 1.20 & 2460518.6550 & 90.59 & 2.20 \\
2460500.9133 & -20.99 & 1.80 & 2460574.5365 & -48.82 & 0.91 & 2460531.6876 & 2.83 & 4.80 \\
2460543.8038 & -27.22 & 1.30 & 2460574.5365 & 43.80 & 1.00 & 2460545.6633 & -75.33 & 3.20 \\
2460545.7649 & -10.76 & 1.00 & \multicolumn{3}{c}{TIC 302581695 (OES)} & 2460545.6633 & 86.48 & 3.80 \\
2460560.7470 & 3.02 & 0.67 & 2460428.3639 & 0.00 & 0.00 & 2460556.6051 & 4.77 & 0.93 \\
2460565.7252 & -8.31 & 1.40 & 2460429.4146 & -11.76 & 1.50 & 2460568.7067 & 6.72 & 1.30 \\
2460567.8044 & -27.52 & 1.10 & 2460430.3624 & 1.47 & 1.20 & \multicolumn{3}{c}{TIC 419610625 (PUCHEROS+)} \\
2460568.7751 & -0.10 & 0.01 & 2460431.3573 & -15.86 & 1.40 & 2460499.8637 & -24.37 & 0.24 \\
2460575.6070 & -27.02 & 2.10 & 2460433.3454 & -15.03 & 1.50 & 2460500.8190 & 15.06 & 0.31 \\
\multicolumn{3}{c}{TIC 226037840 (PUCHEROS+)} & 2460435.3046 & -17.48 & 2.10 & 2460510.6786 & -22.14 & 0.35 \\
2460517.5634 & -11.30 & 7.20 & 2460436.3112 & 1.18 & 2.10 & 2460518.6633 & -6.01 & 0.27 \\
2460517.5831 & -30.85 & 16.00 & 2460707.5776 & -14.95 & 1.90 & 2460519.7562 & -13.00 & 0.36 \\
2460519.6531 & -2.80 & 6.00 & 2460709.3922 & -18.26 & 2.00 & 2460543.7860 & -7.29 & 0.36 \\
2460545.5686 & -2.67 & 7.80 & 2460711.2407 & -16.08 & 1.70 & 2460545.7477 & 22.26 & 0.51 \\
2460569.5620 & -6.98 & 5.80 & 2460715.3252 & -20.93 & 2.40 & 2460559.6020 & 21.31 & 0.27 \\
\multicolumn{3}{c}{TIC 257456854 (OES)} & 2460722.5093 & 2.28 & 1.90 & 2460565.6900 & 12.13 & 0.32 \\
2460428.5319 & -79.31 & 0.56 & \multicolumn{3}{c}{TIC 302581695 (MUSICOS)} & 2460568.7185 & -0.04 & 0.00 \\
2460428.5319 & 60.97 & 0.62 & 2460712.2732 & -2.66 & 4.80 & \multicolumn{3}{c}{TIC 444577764 (PUCHEROS+)} \\
2460430.5294 & -62.31 & 0.37 & 2460715.4068 & -23.48 & 4.30 & 2460554.5447 & -15.75 & 6.00 \\
2460430.5294 & 44.05 & 0.46 & 2460717.5150 & -23.09 & 3.60 & 2460564.5294 & -0.08 & 0.01 \\
2460431.4949 & -38.80 & 0.50 & 2460728.4267 & -2.92 & 4.70 & \multicolumn{3}{c}{TIC 448876509 (PUCHEROS+)} \\
2460431.4949 & 16.13 & 0.45 & \multicolumn{3}{c}{TIC 302666414 (OES)} & 2460516.5070 & 0.00 & 0.00 \\
2460433.4819 & -31.81 & 0.43 & 2460431.4031 & 0.00 & 0.00 & 2460518.5273 & -2.67 & 0.20 \\
2460433.4819 & 7.93 & 0.49 & 2460433.3905 & -28.27 & 0.15 & 2460530.4900 & -1.63 & 0.39 \\
2460511.4087 & -11.78 & 0.26 & 2460435.3505 & -14.62 & 0.35 & 2460564.5795 & -2.26 & 0.62 \\
2460512.4158 & -37.99 & 0.44 & 2460436.3573 & -32.27 & 0.22 & 2460569.4989 & -1.21 & 0.57 \\
2460512.4158 & 15.77 & 0.55 & 2460709.4763 & -24.73 & 0.10 & 2460570.5614 & -2.18 & 0.26 \\
2460520.3557 & -33.48 & 0.42 & 2460715.4836 & -32.42 & 0.20 &  &  &  \\
2460520.3557 & 8.61 & 0.42 & 2460716.5513 & -6.25 & 0.21 &  &  &  \\
2460523.3679 & -75.56 & 0.59 & 2460722.3418 & -24.76 & 0.43 &  &  &  \\
2460523.3679 & 57.68 & 0.65 & 2460722.4848 & -19.47 & 0.20 &  &  &  \\
2460526.3847 & -60.77 & 0.46 & \multicolumn{3}{c}{TIC 302666414 (MUSICOS)} &  &  &  \\
2460526.3847 & 34.97 & 0.42 & 2460712.3473 & -31.48 & 0.43 &  &  &  \\
2460559.2909 & -78.37 & 0.52 & 2460715.5680 & -31.76 & 0.46 &  &  &  \\
2460559.2909 & 58.25 & 0.61 & 2460717.5401 & -9.42 & 0.40 &  &  &  \\
\multicolumn{3}{c}{TIC 257456854 (MUSICOS)} & 2460728.5046 & -29.92 & 0.32 &  &  &  \\
2460830.4519 & -50.01 & 0.43 & \multicolumn{3}{c}{TIC 310932102 (OES)} &  &  &  \\
2460830.4519 & 27.37 & 0.54 & 2460511.4820 & -0.99 & 0.85 &  &  &  \\
2460833.4517 & -80.00 & 0.53 & 2460512.4608 & -2.39 & 0.70 &  &  &  \\
2460833.4517 & 53.69 & 0.49 & 2460520.3929 & -3.28 & 0.84 &  &  &  \\
2460835.3559 & -40.75 & 0.51 & 2460523.4049 & -0.02 & 0.00 &  &  &  \\
2460835.3559 & 15.55 & 0.47 & 2460856.4279 & 3.32 & 1.00 &  &  &  \\
2460836.3840 & -23.39 & 0.94 & 2460870.3612 & 5.39 & 0.89 &  &  &  \\
2460836.3840 & -4.65 & 0.49 & \multicolumn{3}{c}{TIC 310932102 (MUSICOS)} &  &  &  \\
2460838.3924 & -47.68 & 0.52 & 2460860.3476 & 2.81 & 3.00 &  &  &  \\
2460838.3924 & 21.47 & 0.52 & \multicolumn{3}{c}{TIC 320692159 (MUSICOS)} &  &  &  \\
2460848.4524 & -55.76 & 1.00 & 2460860.3842 & -28.37 & 4.20 &  &  &  \\
2460848.4524 & 34.51 & 1.10 & 2460860.3842 & 37.30 & 5.30 &  &  &  \\
\multicolumn{3}{c}{TIC 279821618 (PUCHEROS+)} & 2460861.3828 & -36.30 & 4.80 &  &  &  \\
2460517.5423 & 4.12 & 2.90 & 2460861.3828 & 45.81 & 6.10 &  &  &  \\
2460519.6262 & 6.55 & 3.50 & \multicolumn{3}{c}{TIC 342829903 (PUCHEROS+)} &  &  &  \\
2460530.5613 & 8.65 & 4.50 & 2460516.4869 & -84.93 & 2.40 &  &  &  \\
2460545.5437 & 4.41 & 6.10 & 2460516.4869 & 101.44 & 2.50 &  &  &  \\
2460547.5831 & 6.48 & 5.60 & 2460518.4858 & 6.71 & 1.40 &  &  &  \\
2460556.5188 & 0.05 & 0.03 & 2460522.5411 & -84.09 & 7.30 &  &  &  \\
2460565.5278 & 2.29 & 3.70 & 2460522.5411 & 105.90 & 7.60 &  &  &  \\
2460569.5407 & 2.71 & 4.20 & 2460530.4700 & -85.81 & 2.40 &  &  &  \\
\multicolumn{3}{c}{TIC 292207311 (PUCHEROS+)} & 2460530.4700 & 102.88 & 2.40 &  &  &  \\
2460517.4984 & 7.60 & 1.70 & \multicolumn{3}{c}{TIC 347699402 (OES)} &  &  &  \\
2460519.5324 & -6.27 & 1.70 & 2460511.4242 & 0.29 & 0.13 &  &  &  \\
2460530.5109 & -4.66 & 2.00 & 2460512.4345 & 0.12 & 0.24 &  &  &  \\
2460558.4734 & -3.52 & 1.50 & 2460520.3714 & -0.89 & 0.07 &  &  &  \\
2460566.5308 & 5.15 & 3.10 & 2460523.3835 & 0.97 & 0.29 &  &  &  \\
2460577.5024 & -0.03 & 0.01 & 2460526.4005 & -0.06 & 0.06 &  &  &  \\
\multicolumn{3}{c}{TIC 293069615 (PUCHEROS+)} & 2460526.4171 & 0.02 & 0.00 &  &  &  \\
2460514.5269 & -84.33 & 1.40 & 2460870.5444 & 0.38 & 0.12 &  &  &  \\
2460514.5269 & 80.98 & 1.20 & \multicolumn{3}{c}{TIC 347699402 (MUSICOS)} &  &  &  \\
2460519.6869 & -84.33 & 1.20 & 2460859.4025 & -0.33 & 0.72 &  &  & \\ \hline \hline
\end{tabular}
\end{table}

\end{appendix}

\end{document}